\documentclass[%
 reprint,
superscriptaddress,
 amsmath,amssymb,
 aps,
]{revtex4-2}

\usepackage{graphicx}% Include figure files
\usepackage{dcolumn}% Align table columns on decimal point
\usepackage{bm}% bold math
\usepackage{bbm}
\usepackage{amsmath}
\usepackage{braket}
\usepackage{xcolor,verbatim}
\usepackage{mathtools}
\usepackage[normalem]{ulem}

\begin{document}

\preprint{APS/123-QED}

\title{Analysis of satellite-to-ground quantum key distribution with adaptive optics}
%\author{}
%\date{June 2020}


\author{Valentina Marulanda Acosta}
 \email{valentina.marulanda-acosta@lip6.fr}
 \affiliation{DOTA, ONERA, Université Paris Saclay, F-92322 Châtillon, France}
 \affiliation{Sorbonne Université, CNRS, LIP6, F-75005 Paris, France}
\author{Daniele Dequal}
 \email{daniele.dequal@asi.it}
 \affiliation{Telecommunication and Navigation Division, Agenzia Spaziale Italiana, Matera, Italy}
\author{Matteo Schiavon}
 \affiliation{Sorbonne Université, CNRS, LIP6, F-75005 Paris, France}
\author{Aurélie Montmerle-Bonnefois}
 \affiliation{DOTA, ONERA, Université Paris Saclay, F-92322 Châtillon, France}
\author{Caroline B. Lim}
 \affiliation{DOTA, ONERA, Université Paris Saclay, F-92322 Châtillon, France}
\author{Jean-Marc Conan}
 \affiliation{DOTA, ONERA, Université Paris Saclay, F-92322 Châtillon, France}
\author{Eleni Diamanti}
 \affiliation{Sorbonne Université, CNRS, LIP6, F-75005 Paris, France}

\date{\today}% It is always \today, today,

\begin{abstract}
Future quantum communication infrastructures will rely on both terrestrial and space-based links integrating high-performance optical systems engineered for this purpose. In space-based downlinks in particular, the loss budget and the variations in the signal propagation due to atmospheric turbulence effects impose a careful optimization of the coupling of light in single-mode fibers required for interfacing with the receiving stations and the ground networks. In this work, we perform a comprehensive study of the role of adaptive optics (AO) in this optimization, focusing on realistic baseline configurations of prepare-and-measure quantum key distribution (QKD), with both discrete and continuous-variable encoding, and including finite-size effects. Our analysis uses existing experimental turbulence datasets at both day and night time to model the coupled signal statistics following a wavefront distortion correction with AO, and allows us to estimate the secret key rate for a range of critical parameters, such as turbulence strength, satellite altitude and ground telescope diameter. The results we derive illustrate the interest of adopting advanced AO techniques in several practical configurations.
\end{abstract}


\maketitle

\section{\label{sec:level1}Introduction}

The growing needs for high data rate confidential communications and the security threats posed by the forthcoming advent of quantum computers make it critical to develop a higher security standard for our future data transmissions.
This can be achieved with quantum key distribution (QKD), a communication paradigm that exploits the quantum properties of a single photon in a superposition of a finite number of optical modes (discrete variables, DV) or of the quadratures of low intensity coherent beams (continuous variables, CV) to guarantee a higher level of security ~\cite{Scarani2009}. 
%Ralph2000, Grosshans2003,
While QKD systems have reached technological  maturity \cite{diamanti_practical_2016}, due to the exponential attenuation of light in optical fibers, terrestrial links are limited to few hundred kilometers distances without the deployment of quantum repeaters ~\cite{chen2020, Chen2021, Pittaluga2021}. For this reason, satellite-to-ground links appear as a promising alternative to implement intercontinental communication \cite{liao_satellite-relayed_2018}.

The feasibility of free-space and satellite QKD has been the topic of numerous studies, both theoretical and experimental, over the years \cite{Vasylyev2012, vasylyev_satellite-mediated_2019,pirandola_advances_2020,shen_free-space_2019}. In particular, considerable effort has been devoted to the estimation of the secret key rate achievable over a satellite link \cite{Liorni2019,Polnik2020, dequal_feasibility_2021, Ecker2021}. This is due to the fact that, unlike fiber connections, satellite links are subject to several specific effects, such as the impact of atmospheric turbulence on beam broadening and wandering \cite{wang_atmospheric_2018,ruppert_fading_2019, kish_feasibility_2020,pirandola_satellite_2021,zuo_atmospheric_2020,dequal_feasibility_2021}, the effect of divergence and pointing error \cite{Vasylyev2012, vasylyev_satellite-mediated_2019}, the wavelength shift due to Doppler effect and the finite duration of the key exchange, when the orbit considered is not geostationary \cite{sidhu2021finite}.

%Recent work regarding the impact of atmospheric channel on the performance of CV protocols can be found in \cite{wang_atmospheric_2018,ruppert_fading_2019, kish_feasibility_2020,pirandola_satellite_2021,zuo_atmospheric_2020,dequal_feasibility_2021}. 

%In their article, Dequal et al. \cite{dequal_feasibility_2021} account for the impact of beam wandering on the secret key rate in the case of continuous variables (CV) QKD, and assume as a first approximation that the atmospheric turbulence-induced fluctuations of the transmission efficiency be perfectly compensated. The transmission efficiency through atmospheric turbulence was considered constant equal to 50\%.

Besides these effects, the impact of turbulence on the spatial coherence of the beam, and as a result on the ability to couple the propagating beam into a single mode component for detection, can be disastrous. Such coupling is typically necessary when using coherent detectors for CV-QKD or superconducting nanowire single-photon detectors (SNSPD) for DV-QKD.
 % and the impact of a turbulent free-space channel was studied in \cite{liorni_satellite-based_2019}, as well as in \cite{jin_genuine_2019} for time-bin based protocols or in \cite{sit_high-dimensional_2017} for orbital angular momentum based protocols.  \\
% A first experimental demonstration using a CV protocol has been recently proposed \cite{shen_free-space_2019}. The 460 m link was based on a uni-dimensional polarization encoding, and the local oscillator was sent on one of the two polarization directions to avoid using adaptive optics (AO). 
%Regarding DV protocols, much progress was made with the experimental demonstrations on the Micius payload \cite{liao_satellite--ground_2017,liao_satellite-relayed_2018, yin_satellite-based_2017} using polarization encoding, and the impact of a turbulent free-space channel was studied in \cite{liorni_satellite-based_2019}, as well as in \cite{jin_genuine_2019} for time-bin based protocols or in \cite{sit_high-dimensional_2017} for orbital angular momentum based protocols.  \\
The use of adaptive optics (AO) to compensate for the free-space channel induced perturbations is a promising solution. Because they offer the possibility to compensate in real time the phase fluctuations induced by atmospheric turbulence, AO systems are commonly used in astronomy \cite{ciliegi_maory_2020} and are becoming a key technology for free-space optical telecommunications \cite{chen_performance_2018}, as well as for free-space QKD \cite{gruneisen_adaptive-optics-enabled_2021, gruneisen_adaptive_2014, gruneisen_modeling_2017,oliker_how_2019,chai_suppressing_2020,wang_performance_2019,pugh_adaptive_2020}.
Further effort thus needs to be made to finely model the atmospheric channel, in view of proposing a dedicated means of optimizing the link performance and of ultimately reaching a reliable system design  \cite{wright_adaptive_2015,petit_investigation_2016}.

In this work, we consider a link from a Low Earth Orbit (LEO) satellite to an optical ground station (OGS). We account for the impact of atmospheric turbulence on the transmission efficiency probability distribution as well as the effect of an AO system custom designed to compensate for this impact. We then assess the feasibility of employing such a link to distribute a secret key. The overall link performance is hence characterized in terms of secret key rate for a satellite-to-ground downlink scenario. We consider different AO configurations of increasing complexity, assuming various turbulence conditions, at both day and night time.
We also consider both DV and CV-QKD protocols for typical scenarios that however correspond in general to different maturity level for the involved technologies and hence should not be directly compared.

%however our work is limited to estimations of key rates in what could be considered as typical scenarios and it excludes any comparison of both protocols, due to different technologies at different readiness levels being involved.

In the following, first we describe the propagation channel model and chosen turbulence scenarios and estimate the probability distribution of transmission efficiency after AO correction, based on a pseudo-analytical simulation tool. Subsequently, we describe our method and parameters for estimating the secret key rates for DV and CV-QKD. Finally, we discuss the obtained results in terms of secret key rate and assess their sensitivity to system parameters like the satellite altitude, the ground telescope diameter, and the turbulence strength.

% Underline the importance of AO for single mode coupling (DV, superconducting detectors) and wave-front correction (CV).During its propagation through the atmospheric channel, the spatial coherence of the beam can be severely degraded, which may be critical for single mode coupling into a single mode fiber or into a  superconducting nanowire single photon detector (SNSPD). With the possibility to compensate in real-time the phase fluctuations induced by atmospheric turbulence, AO systems are commonly used in astronomy and are becoming a key technology for free space optical telecommunication systems. \\

%Analysis performed with high performance AO and "basic" AO.

%{\color{red} For CV: do we loose mode-matched signal in single mode coupling? Would it be better to compensate the LO instead of the quantum signal? For DV: are there superconducting detectors with multimode fiber?  General remark: single mode coupling filters out noise. Would it be interesting to see the impact at different wavelengths (850 nm, 532nm), quantifying the noise filtering property of AO? Maybe it might be possible to do daylight QKD also in the visible, with strong filtering... } \\

{\color{red} %Prelim note for writing: The objective of this paper is the key rate estimation of satellite QKD for downlink in daylight and nighttime scenarios, considering both DV and CV and 1550 nm wavelength.\\
%Reference to previous work accounting for beam wandering and average signal attenuation due to beam propagation. Following up accounting for turbulence-induced phase distortion and resulting signal fluctuations.\\
}

\section{Propagation channel model}

We consider a downlink scenario where a LEO satellite establishes a secret key with an optical ground station. We assume a circular orbit passing at the ground station zenith. Our analysis here is limited to elevation angles above 20 degrees for which phase and amplitude perturbations induced by turbulence remain compatible with state of the art adaptive optics systems. 

In our baseline scenario, the satellite emits a beam with a divergence of $\theta_d~=~10 \,\mu$rad and points it towards a ground station telescope with a diameter of 1.5~m. The propagation is affected by the wandering of the beam mainly due to the pointing error of the satellite. This error is characterized by its standard deviation taken here equal to $\theta_p=1\, \mu$rad. In addition to that, the atmospheric turbulence present on the propagation channel induces phase and amplitude fluctuations on the beam. Since the effects of the satellite pointing error and of the atmospheric turbulence are independent, in this work we assume that they can be treated separately.

\subsection{\label{sec:level2}Atmospheric model}

To account for atmospheric turbulence induced effects, some assumptions need to be made regarding the turbulence profile along the line of sight between the ground station and the satellite. The turbulence strength is characterized locally along a certain path by $C_n^2$, the so-called refractive-index structure constant. To account for the total strength of turbulence along the line of sight, we use integrated parameters: the Fried parameter $r_0$, which represents the equivalent aperture diameter setting the telescope angular resolution limit in the presence of turbulence, the correlation time $\tau_0$, which characterizes the temporal variation of the wavefront, the isoplanatic angle $\theta_0$, which is the angle between two propagation paths for which the wavefront distortion can be considered identical, and the log-amplitude scintillation variance $\sigma^2_\chi$, which represents the fluctuation of the flux in the receiving pupil caused by the impact of turbulence on the propagated wave.

For a plane wave propagation at a given distance and turbulence statistics described by a Kolmogorov spectrum, we have:
\begin{eqnarray}
r_0 = \left[0.423  \left(\frac{2\pi}{\lambda}\right)^2 \right. 
\left. \int_{0}^{z_{max}} C_n^2(z) \,dz \right]^{-3/5}.
\end{eqnarray}
Under the frozen flow hypothesis, turbulent flows are frozen and will move across the pupil at wind speed $V(z)$; we then have:
\begin{eqnarray}
\tau_0 = \left[ 2.91 
\left(\frac{2\pi}{\lambda}\right)^2 
\int_{0}^{z_{max}} V(z)^{5/3}C_n^2(z) \,dz
\right]^{-3/5}.
\end{eqnarray}
Besides: 
\begin{eqnarray}
\theta_0 = \left[ 2.91  \left(\frac{2\pi}{\lambda}\right)^2 
\int_{0}^{z_{max}} z^{5/3}C_n^2(z) \,dz
\right]^{-3/5}.
\end{eqnarray}
Given that we consider elevations above $20^\circ$ we work under a weak perturbation regime and thus:
\begin{eqnarray}
\sigma^2_{\chi} = 
0.5631
\left(\frac{2\pi}{\lambda}\right)^{7/6}
\int_{0}^{z_{max}} z^{5/6}C_n^2(z) \,dz.
\end{eqnarray}
In the above expressions, $z$ is the distance from the receiving pupil along the line of sight and $z_{max}$ corresponds to the distance to the uppermost layer of the turbulent volume where the value of  $C_n^2$ becomes negligible. $\lambda$ is the wavelength and $V(z)$ corresponds to the wind speed which in the case of a LEO satellite includes both the natural wind and the apparent wind induced by the satellite moving across the sky. The latter is defined as $V_{app}(z) = \dot{\theta}z$ with $\dot{\theta}$ the slew rate of the satellite.

Most of the existing turbulence databases correspond to nighttime measurements taken at astronomical sites. However, our QKD link should possibly be available during the day as well. We have therefore decided to construct some statistically representative $C_n^2$ profiles based on different sets of measurements as described in \cite{vedrenne_performance_2021}. We start from the hypothesis that the integrated parameters $\theta_0$ and $r_0$ are mostly independent from one another. The former is determined mainly by the atmosphere at high altitudes (above 2000~m), while the latter depends mostly on the lower layers of the atmosphere.

The $C_n^2$ values for the upper layers for both day and night profiles are extracted from an existing set of measurements taken at Cerro Paranal \cite{osborn_optical_2018}. The statistical distribution of the isoplanatic angle $\theta_0$ is computed from the entire set and a value is selected. The corresponding $C_n^2$ values constitute the first part of the turbulence profile. The low atmosphere $C_n^2$ values are derived from two different Canary Islands databases; seeing values for nighttime  \cite{vazquez_ramio_european_2012} and $C_n^2$ measurements at 30~m for daytime \cite{sprung_characterization_2013}. The behavior of $C_n^2$ for the low layers is deduced using a Monin-Obhukov similitude law which describes the evolution of this parameter on the surface layer as a function of height. These values permit the computation of the statistical distribution of the Fried parameter $r_0$ and a value is then chosen according to the intensity of the turbulence we wish to represent. The values obtained for high and low layers are then brought together, resulting in the final hybrid turbulence profile.

\begin{figure}[h]
\centering
%\begin{subfigure}%{.5\textwidth}
 % \centering
  \includegraphics[width=80mm]{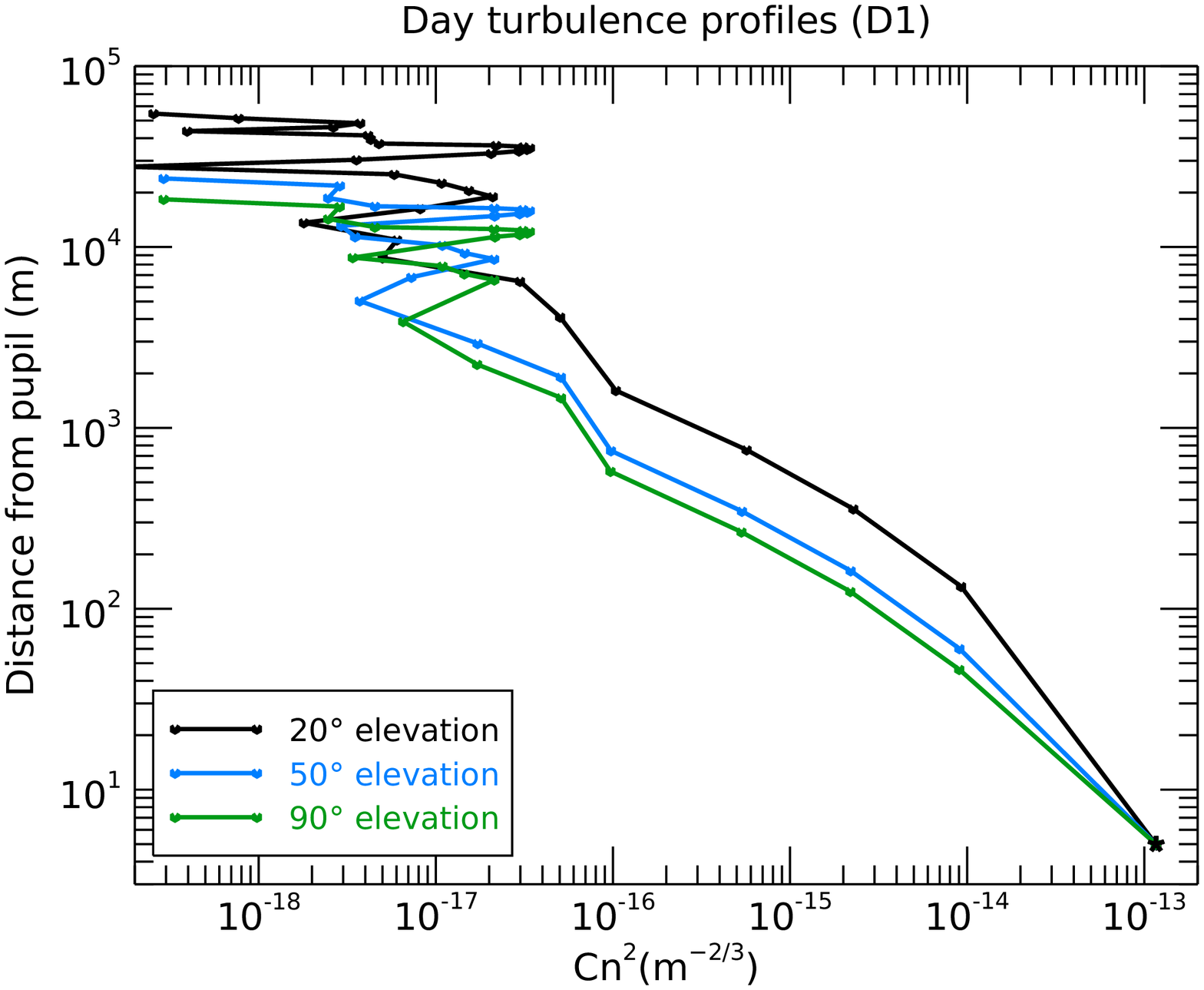}
%\end{subfigure}%
%\begin{subfigure}%{.5\textwidth}
 % \centering
  \includegraphics[width=80mm]{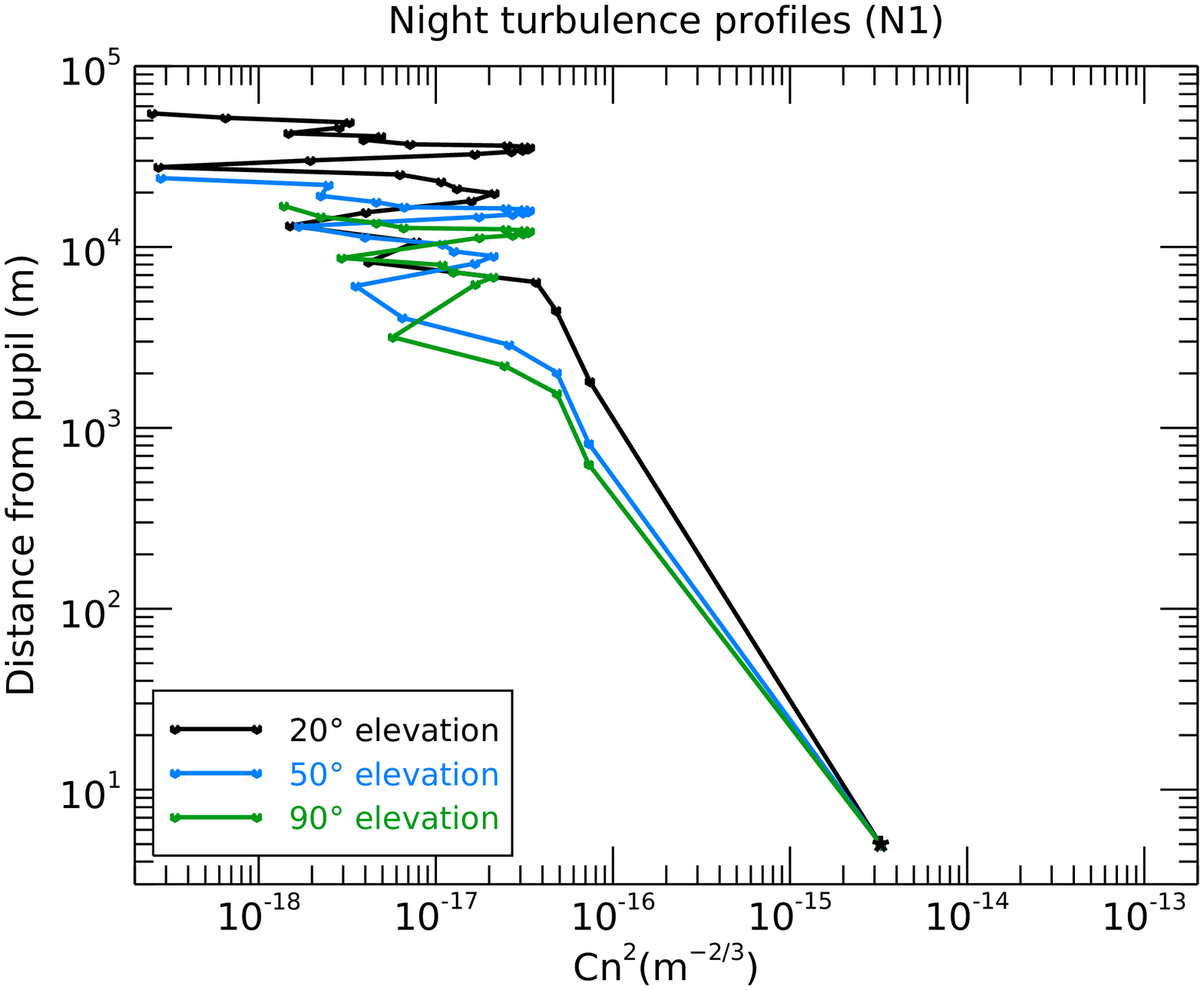}
%\end{subfigure}
\caption{Median turbulence profiles D1 (top) and N1 (bottom).}
\label{profiles}
\end{figure}

As a baseline, we choose day and night profiles corresponding to median turbulence conditions in terms of $\theta_0$ and $r_0$ values \cite{vedrenne_performance_2021}; we will refer to them in the following as D1 and N1. Even if our specific study case does not have to deal with the effect of anisoplanatism, the $\theta_0$ value is important because of its link with $\tau_0$ as $\tau_0 \approx \theta_0/\dot{\theta}$. Figure~\ref{profiles} shows some examples of the resulting profiles and Table \ref{int_parameters} sums up the corresponding values of the turbulence integrated parameters at the highest and lowest elevations considered and for a wavelength of 1550~nm. 

    \begin{table}[h]
        \centering
         \begin{ruledtabular}
        \begin{tabular}{ccccc}
    \hline
    &\multicolumn{2}{c}{D1}&\multicolumn{2}{c}{N1}\\
    \hline
    & $20^\circ$ &$90^\circ$&$20^\circ$&$90^\circ$\\
     \hline
      %$r_0$ (cm) & 7.86 & 14.96 & 26.55 &50.44\\
      $r_0$ (cm) & 7.9 & 15.0 & 26.6 &50.4\\
     \hline
     $\tau_0$ (ms) &1.7 &1.8 &1.7 & 1.8\\     \hline
      %$\theta_0$ ($\mu$rad) & 6.218  &34.554 &6.231&34.463\\
      $\theta_0$ ($\mu$rad) & 6.2  &34.5 &6.2&34.4\\
     \hline
      %$\sigma_\chi ^2$ & 0.0692  & 0.0098 & 0.0583& 0.0083\\
       $\sigma_\chi ^2$ & 0.07  & 0.01 & 0.05& 0.008\\
     \hline
    \end{tabular}
    \end{ruledtabular}
    \caption{Integrated parameters for the baseline day and night profiles.}
    \label{int_parameters}
\end{table}

\subsection{Adaptive optics system}
The propagation of the quantum signal through the atmospheric turbulence affects significantly both its amplitude and its phase, causing aberrations that impact negatively the detection process. The use of an AO system allows for an improvement of the coupling efficiency of the incoming wave into the core of a single mode optical fiber (SMF), which will result in a better recovery of the quantum signal and ultimately an increase on the key rate. It is worth underlying that the SMF is used as spatial mode filter for CV-QKD, allowing for a high visibility interference with the local oscillator of the coherent detector. For DV-QKD, SMF is used both as a spatial filter to suppress stray light, especially in daylight scenario \cite{Avesani2021}, and as a way to efficiently couple the incoming light to high-efficiency, low-jitter SNSPD.

\begin{figure}[h]
\centering
 \includegraphics[width=80mm]{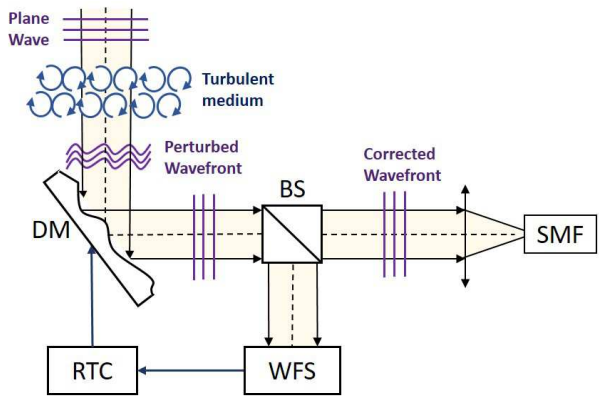}
\caption{General scheme of an adaptive optics system. WFS: Wavefront sensor, DM: Deformable mirror, RTC: Real time controller, BS: beam splitter, SMF: Single mode fiber}.
\label{ao_scheme}
\end{figure}

A typical AO system is shown in Fig.~\ref{ao_scheme}. After a plane wave is perturbed by a turbulent medium it enters a closed control loop consisting of three main components. The wave is first reflected on a deformable mirror (DM) and the beam splitter redirects a portion of it to a wavefront sensor (WFS) which detects the aberrations of the signal. Then, a real time controller (RTC) gives instructions to the DM in order to adapt to the aberrations. The beam resulting from this correction is then coupled into a single mode fiber. This type of system has already proven to be effective in improving the quality of the received signal for classical communication applications \cite{petit_investigation_2016}. Since the QKD signal is at the single or low photon level, an additional beacon at a slightly different wavelength will have to be used as a probe to measure and correct the phase aberrations of the wavefront. Because of the achromatic nature of turbulence, the wavelength difference is not an issue when estimating the aberrations, inducing at worst an additional error term that will not be considered within the scope of this work.

As a baseline, we consider an AO system with a 2 frame delay loop at a sampling frequency of 5 kHz. The aberrations of the turbulent wavefront will be modeled by decomposing them into 861 Zernike modes (equivalent to 40 radial orders \cite{noll_zernike_1976}).
Using the pseudo-analytical modeling tool SAOST \cite{hemmati_adaptive_2016,canuet_statistical_2018,conan_adaptive_2019}, we can estimate, for a given scenario of satellite orbit, turbulence profile and AO system parameters, the performance of the AO correction in terms of scintillation effects and phase aberration residuals as well as the statistical properties of the SMF coupling efficiency of the corrected beam. SAOST employs a plane wave hypothesis and a weak fluctuation regime approximation, \emph{i.e.}, typical log-amplitude scintillation variance values $\sigma^2_\chi < 0.3$. These are suitable assumptions for our scenario taking into account the fact that the beam size at the receiver is significantly larger than the telescope and that we only consider elevations above $20^\circ$. The coupling efficiency of the turbulent wavefront after correction is computed as the magnitude squared of the overlap integral between the incoming corrected complex amplitude and the main mode of the fiber (or of the local oscillator laser). The ratio between the receiving aperture diameter and the beam waist of the fiber mode at the focal plane is taken as: $D_{RX}/w_0 = 2.2$.

The tool allows us to consider different correction configurations, elevation and turbulence intensities in order to determine the correction level best suited to our particular application. Indeed, the AO system design shall be the result of a trade-off between system cost and complexity and ultimate performance. This analysis is shown later in Section IV, where we assess the secret key rate performance for AO systems able to correct aberrations up to 5, 10, 15 and 20 Zernike radial orders.

Assuming a D1 profile, Table~\ref{Error budget} shows the variance of the turbulent phase aberrations before correction, and the error budget of an AO system correcting 15 radial orders, for the lowest and highest elevations considered in this study. Indeed, adaptive optics correction has its limitations and there will always be a remaining amount of phase aberrations the system is not able to correct. The initial turbulent phase aberrations are higher for lower elevations, which is expected, and the residual error after AO correction originates from the different components of the AO. The fitting error is the most prominent and corresponds to the phase aberrations that could not be corrected because they lie beyond the number of radial orders corrected by the system. Increasing the complexity of the deformable mirror (\emph{i.e.}, the number of actuators) would enable to increase the number of corrected radial orders, in order to reduce the fitting error to a target value. The aliasing error is due to the finite number of measurements performed by the wavefront sensor when sampling the signal; its value is set to around 35$\%$ of the fitting error. The temporal error is linked to the 2 frame delay of our closed control loop; its low value below $0.1$~rad$^2$ indicates that a 5 kHz frequency is acceptable for this system. In the case of a nighttime profile with a similar correction system, the error variances involved would be smaller in general due to a milder turbulence intensity. 

\begin{table}[h]
\centering
\begin{ruledtabular}
\begin{tabular}{ccc}
\hline
%& & \multicolumn{2}{Variance ($rad^2$)}
 \textbf{Elevation} &\textbf{$20^\circ$ }&\textbf{$90^\circ$}  \\ \hline \hline
\textbf{Turbulent phase} (rad$^2$) & \textbf{140.22%14
}&  \textbf{47.99
%}&  \textbf{47.98%74
}\\\hline \hline
Fitting error (rad$^2$) & 0.46%10 
& 0.16
%& 0.15%77
\\\hline
Aliasing error (rad$^2$) & 0.16%13
& 0.05%52
\\\hline
Temporal error (rad$^2$) & 0.08
%Temporal error ($rad^2$) & 0.07%88
& 0.07
%& 0.06%84
\\\hline
\textbf{Total residual error} (rad$^2$) & \textbf{0.70%13
}& \textbf{0.28%14
}\\\hline
\end{tabular}
\end{ruledtabular}
\caption{Error budget for a daytime link at 400~km, assuming turbulence profile D1 and 15 corrected radial orders.}
\label{Error budget}
\end{table}

The phase aberrations inflicted on the optical signal are dependent on the elevation angle and therefore the performance of an AO system will not be constant throughout the orbit. Previous studies such as \cite{wright_adaptive_2015,petit_investigation_2016} have already implemented experimental LEO-to-ground downlink configurations that permit the analysis of the performance of an AO system as well as of its limitations in practical settings. This performance provides an improvement in signal quality as expected from the theoretical analysis and simulations. These studies, however, are carried out in the context of classical telecommunications and the differences with our scenario need to be taken into account. Two such important differences is the use here of a bigger receiving telescopes, namely of 1.5~m diameter instead of the few tens of cm typical of the telecom case, and the performance metric, since here we do not look to optimize data rates but instead we focus on obtaining a high secret key generation rate.

\subsection{Coupled signal statistics}
\label{PDTE_calc}
We will now consider of all the contributions to the channel attenuation, which will allow us to calculate the overall probability distribution of the transmission efficiency (PDTE) of the communication channel. As was done in \cite{dequal_feasibility_2021}, we start by modeling the satellite dynamics and deriving the time evolution of the satellite-to-ground distance. In order to reduce the number of free parameters and the complexity of the overall model, as noted earlier we limit the analysis to a circular orbit passing at the zenith of the receiving station. After calculating the time evolution of the satellite distance, we proceed with a temporal subdivision of the overall pass in several segments, each of which can be approximated as a fixed distance transmission. We then consider individually each of these segments and calculate their transmission efficiency statistics. Finally, we merge the contributions from all segments to obtain the overall PDTE for the whole pass. 

The first attenuation phenomenon considered is the geometrical loss, which includes both the effect of the beam broadening, mainly due to diffraction, and the beam displacement with respect to the receiver center, due to the pointing error. For each segment defined in the fist step of the analysis, we model the light propagation considering the communication distance fixed, following the model proposed in \cite{vasylyev_satellite-mediated_2019}, where the beam footprint on the ground is approximated as a Gaussian profile and the transmission efficiency is given by the truncation of the Gaussian profile performed by the receiving aperture. According to this model, the PDTE is given by an approximated analytical expression, the log-negative Weibull distribution. Besides divergence and coupling, the signal is attenuated by absorption and scattering through the atmosphere, resulting in an atmospheric transmission efficiency $\tau_{atm}$ given by:
\begin{equation}
    \tau_{\text{atm}}=\tau_{\text{zen}}^{sec(\theta_{\text{zen}})},
    \label{eq:teff}
\end{equation}
where $\theta_{zen}$ is the zenith angle and $ \tau_{zen}$ the transmission efficiency at zenith \cite{Tomasi2014}.

\begin{figure}[h]
\centering
  \includegraphics[width=8cm]{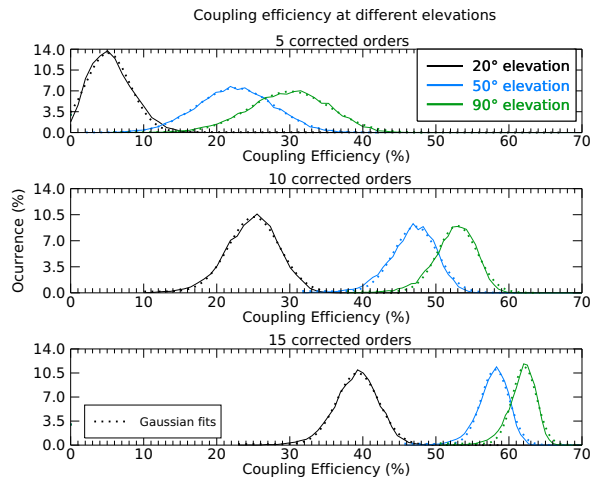}
  \caption{Coupling efficiency into a single mode fiber for different elevations and correction levels considering daytime D1 turbulence conditions.}
\label{ao-coupling}
\end{figure}

The next contribution considered is related to the turbulence-induced wavefront distortion, whose model is described in the previous section. Here we focus on the single-mode fiber (SMF) coupling efficiency, giving a statistical description as a function of different turbulence situations, elevations and AO correction levels. Figure \ref{ao-coupling} shows some examples of the statistical distribution of the SMF coupling efficiency in daylight turbulence conditions for different correction levels and elevation angles. These statistics, derived from the SAOST simulation tool, follow a Gaussian distribution (as evidenced by the Gaussian fits in dotted lines). The coupling efficiency distribution is indeed not constant; it is strongly dependent on the correction capabilities of the AO system, which varies significantly throughout the orbit as the satellite elevation changes. We therefore apply to each orbit segment, performed at the first step of the analysis procedure, the corresponding SMF coupling efficiency statistics. As mentioned during the description of our scenario, geometrical losses and turbulence SMF coupling effects are considered independent and we can then find the joint probability distribution of their product for each orbit segment.

\begin{figure}[h]
\centering
  \includegraphics[width=8cm]{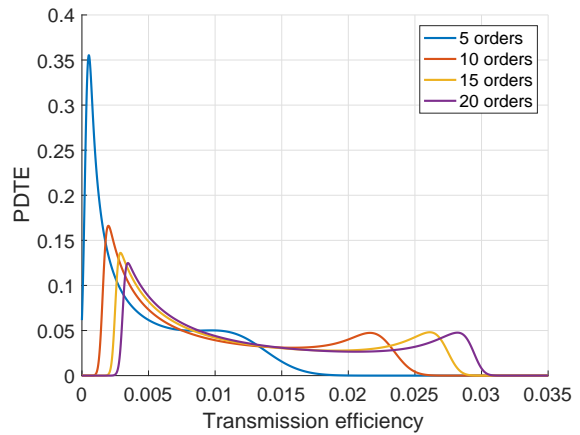}
  \caption{Overall PDTE for a complete satellite pass at 500 km considering daytime D1 turbulence conditions for different AO corrected orders.}
\label{PDTE_vs_AO}
\end{figure}

\begin{figure}[h]
\centering
%\begin{subfigure}%{.5\textwidth}
 % \centering
  \includegraphics[width=8cm]{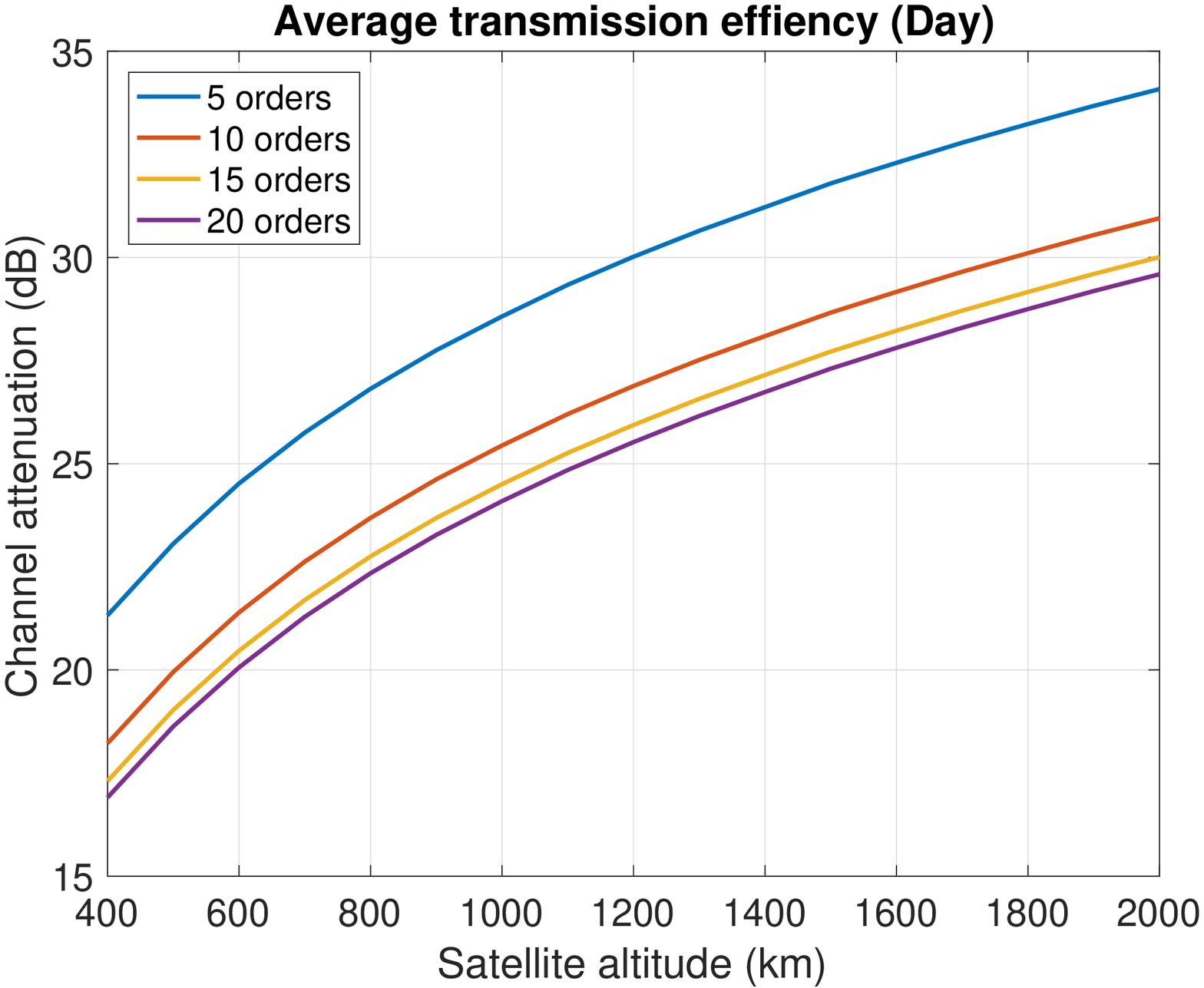}
%\end{subfigure}%
%\begin{subfigure}%{.5\textwidth}
 % \centering
  \includegraphics[width=8cm]{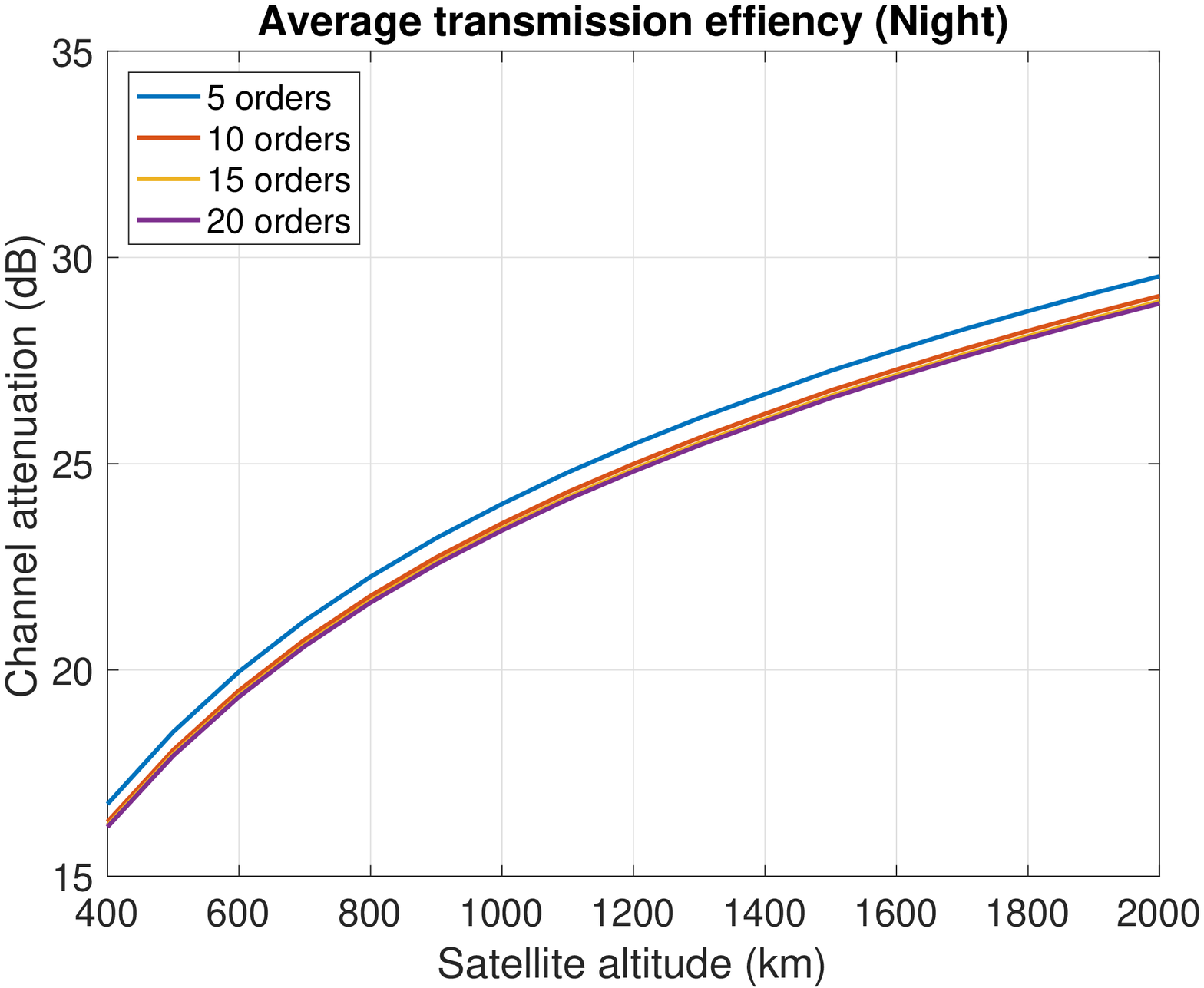}
%\end{subfigure}
\caption{Transmission efficiency for a satellite-to-ground link, averaged over a complete orbit considering daylight D1 (top) and nighttime N1 (bottom) turbulence profiles. }
\label{Teff_avg}
\end{figure}

Finally, we integrate all the contributions related to the orbit segmentation to retrieve the overall geometrical PDTE of the complete satellite pass. As an example, in Fig.~\ref{PDTE_vs_AO}  we report the PDTE for a 500 km altitude satellite with daylight profile D1 for the four AO systems considered in our analysis. In Fig.~\ref{Teff_avg} we can observe the total channel attenuation averaged over the complete satellite pass for different satellite altitudes and correction levels in both day and night median turbulence conditions. The effect of the AO is less evident for nighttime transmission since the turbulence levels, and therefore the fiber coupling losses, are already low in comparison with the daytime scenario. In daytime the channel attenuation is certainly mitigated by the correction of the AO system. This is especially noticeable when increasing the number of radial orders corrected from 5 to 10, which results in a decrease of $\sim$ 3 dB of the average attenuation. While further increasing the number of corrected orders continues to decrease attenuation, the improvement is much less significant for higher correction levels. 

\section{Secret key rate estimation}
\label{key_rate_section}

Let us now describe the basic principles of the discrete and continuous-variable QKD protocols that we consider in this work. We explain in each case the calculation of the estimated secret key generation rate both in the asymptotic regime and taking into account finite-size effects. We then detail the relevant parameters common to both protocols as well as the ones particular to DV or CV-QKD and we provide their respective reference values.

\subsection{Discrete-variable QKD}
%\begin{itemize}
%    \item Efficient BB84 (optimization of base usage)
%    \item  {\color{red} Polarization encoding?}
%    \item Decoy state (optimization of $\mu$)
%    \item Setup specs (dark count, efficiency, intrinsic depolarization, ...)
%    \item Finite size
%\end{itemize}

In DV-QKD, the information is encoded in the superposition of different modes of a single photon (qubits), such as polarization~\cite{Muller1993}, temporal modes~\cite{Townsend1993} or orbital angular momentum~\cite{Vaziri2002}.
The first QKD protocol, proposed by Bennett and Brassard in 1984~\cite{bennett1984} (BB84), exploited two non-orthogonal bases, $Z = \{\ket{0},\ket{1}\}$ and $X = \{\ket{+},\ket{-}\}$, to encode the information.
For each transmitted bit, Alice and Bob choose randomly in which basis to encode and measure the information; then after the transmission, they discard the bits where they used different bases in the so called sifting phase. In this phase, Alice and Bob gain an advantage over a potential Eavesdropper, Eve, whose information can be bounded by looking at the quantum bit error rate (QBER), calculated in a phase called parameter estimation. If the QBER is lower than a certain threshold, it is possible to demonstrate that the mutual information between Alice and Bob is higher than the one shared with Eve, allowing them to distill a secret key using privacy amplification~\cite{Bennett1995}.

The choice of encoding the information in one of the two bases with equal probability reduces the sifted key by a factor $2$, since Bob has $1/2$ probability of choosing the wrong basis.
This can be improved by using the so called efficient BB84~\cite{Lo2004}, where an asymmetry between the two bases is introduced, with one basis, chosen with high probability $q$, used for the actual exchange of the key and the other basis chosen to detect the presence of Eve.
In the asymptotic limit of infinitely long key, it is possible to take $q \sim 1$.

%The nonexistence of efficient single-photon sources~\cite{Eisaman2011} has led to the use of
Typically, attenuated lasers are used to generate qubits for quantum communication protocols~\cite{Scarani2009}. However, an attenuated laser with a mean number of photons per pulse $\mu$ has a probability on the order of $\mu^2$ of emitting a multi-photon pulse. These pulses are intrinsically insecure, since Eve could perform the so called photon number splitting (PNS) attack~\cite{Huttner1995}, where she forwards one photon to Bob and keeps the others in a quantum memory, measuring them after the sifting phase.
To prevent the detrimental effects of this attack on the key rate in high loss channels, the technique of the decoy states has been proposed~\cite{Hwang2003,Lo2005}.
With this technique, Alice randomly changes the statistics of her laser pulses during the transmission, revealing the statistics used for every pulse during the sifting phase. Since Eve must interact with the pulses without knowing the statistics, Alice and Bob can monitor the estimated parameters for the different statistics and calculate better bounds on the information leaked, obtaining positive key rates for longer distances. Furthermore, it has been demonstrated~\cite{Hayashi2007} that a scheme using two decoy states, the vacuum and a weak decoy, gives the same key rates as a scheme with an infinite number of decoy states. For the simulations in the asymptotic limit, we will not go into the details of the decoy scheme, assuming that the parameters are estimated with infinite precision, leaving only the mean number of photons per pulse $\mu$ of the signal state as free parameter. In this limit, the lower bound to the secret key rate is~\cite{Scarani2009,vasylyev_satellite-mediated_2019}:
\begin{align}
\begin{split}
    K &= q \left\{ Q_{\mu,0} + \mathbf{E}[Q_{\mu,1}]\left[ 1 - h\left(\mathbf{E}[\epsilon_{\mu,1}]/\mathbf{E}[Q_{\mu,1}]\right) \right] \right.\\
    &- \left. f_{EC} \mathbf{E}[Q_{\mu}] h\left(\epsilon_{\mu}\right) \right\},
\end{split}
\end{align}
where $\mathbf{E}[\cdot]$ denotes the expected value, $h(x) = -x\log_2(x) - (1-x)\log_2(1-x)$ is the binary Shannon entropy, $Q_{\mu}$ and $\epsilon_{\mu}$ are the gain and the QBER of the signal states, $Q_{\mu,0}$ and $Q_{\mu,1}$ are the gains of, respectively, the vacuum and the single-photon states and $\epsilon_{\mu,1}$ is the QBER of single-photon states.
The QBER of the signal and of the single-photon states are defined, respectively, as $\epsilon_{\mu} = \mathbf{E}[E_{\mu} Q_{\mu}] / \mathbf{E}[Q_{\mu}]$ and $\epsilon_{\mu,1} = \mathbf{E}[E_{\mu,1} Q_{\mu,1}] / \mathbf{E}[Q_{\mu,1}]$.

As we have seen, differently from fiber-based channels, satellite communication links suffer from variations of the transmission efficiency with time, due to atmospheric effects, pointing error and the variation of the link distance due to satellite orbit. For these reasons, the parameters of the channel change with time and it is necessary to take their average over channel fluctuations~\cite{vasylyev_satellite-mediated_2019}.
This effect can be slightly mitigated by grouping the measurements according to the instantaneous channel transmission efficiency, which can be probed by a classical beacon~\cite{liao_satellite--ground_2017,Avesani2021}.
While the effect is probably not as strong as for CV-QKD protocols (see Section \ref{CV} and~\cite{dequal_feasibility_2021}), similar techniques have already been implemented for DV-QKD for fluctuating channels~\cite{Vallone2015}.
The estimation of the parameters from the properties of the channel is detailed in Appendix~\ref{sec:channelProperties}.

% Finite size effects
While the asymptotic secret key rate is important for fixing an upper limit on the performance of a satellite link, a complete analysis of its performance must keep into account finite-size effects.
Our analysis considers an efficient BB84 protocol, with $Z$ and $X$, respectively, the main and the control basis. It also uses a two-decoy scheme, with the vacuum and a weak decoy, and hence the secret key rate is~\cite{Lim2014}:
\begin{equation}
\begin{split}
    K = \frac{1}{N} \Bigg\{ s^L_{Z,0} &+ s^L_{Z,1} \left[ 1 - h(\phi^U_Z) \right] - \lambda_{EC} \\
     &- 6\log_2\frac{21}{\varepsilon_{sec}} - \log_2\frac{2}{\varepsilon_{cor}} \Bigg\},
    \end{split}
\end{equation}
where $s^L_{Z,0}$, $s^L_{Z,1}$ and $\phi^U_Z$ are, respectively, the lower bound on the number of vacuum and single-photon events and the upper bound on the phase error rate associated with single-photon events.
These parameters are estimated from the number of measured events and errors for the different bases and statistics, as described in Appendix~\ref{sec:finiteKeyDV}.
The term $\lambda_{EC}$ represents the number of bits disclosed during error correction~\cite{Lim2014,Tomamichel2017} and $N$ is the total number of bits sent.
The correctness and security parameters~\cite{Lim2014,RENNER2008,Scarani2008}, $\varepsilon_{cor}$ and $\varepsilon_{sec}$, are both  taken as $10^{-10}$.
We run the simulations assuming a source repetition rate of $100$~MHz, optimizing over five parameters: the probability of the $Z$ basis $q$, the mean number of photons per pulse of the signal and weak decoy states, $\mu$ and $\nu$, and their probability $p_{\mu}$ and $p_{\nu}$, with $p_{vacuum} = 1 - p_{\mu} - p_{\nu}$.

\subsection{Continuous-variable QKD\label{CV}}

Differently from DV-QKD, in CV-QKD the encoding is done in continuous degrees of freedom of the quantum states, such as light amplitude and phase, or equivalently light quadratures. As for DV, in CV many protocols have also been proposed, based on different ways to generate and detect the quantum states \cite{diamanti2015}. In this work we consider Gaussian modulation of the light amplitude and phase and heterodyne detection ~\cite{Weedbrook2004}, as for this protocol the security proofs include finite-size effects \cite{Leverrier2015a, Leverrier2017}.  In this protocol, Alice generates a series of coherent states $\ket{\alpha}$, whose amplitude and phase are set according to a Gaussian distribution with variance $V_A$. As $V_A$ is a free parameter, in the following we optimize it, selecting the value that maximises the key rate for each configuration. After the state transmission, Bob detects both quadratures simultaneously via heterodyne detection. In the following analysis, the detector noise is considered to be calibrated, so that it cannot be exploited by Eve. Following this step, Alice and Bob share a common information, which can be expressed as:
\begin{align}
\label{eq_IAB}
\beta I_{AB} = \beta \log_2 \left( 1+ \frac{T^2 V_A}{\sigma^2} \right) ,
\end{align}
where $I_{AB}$ is Alice and Bob's mutual information, $\beta$ is the reconciliation efficiency, $T^2$ is the channel transmission efficiency and $\sigma^2$ is the variance of the noise. 

To set an upper bound on the information available to Eve, Alice and Bob estimate the covariance matrix $\Gamma$ of their data. From the symplectic eigenvalues $\nu_i$ of the covariance matrix, it is then possible to estimate the maximum information available to Eve as:
\begin{align}
\label{eq_chiBE}
\chi_{BE} = g(\nu_1) + g(\nu_2) - g(\nu_3) - g(\nu_4),
\end{align}
where $g$ is the entropy function, $g(z)=\frac{z+1}{2} \log_2\frac{z+1}{2} - \frac{z-1}{2} \log_2\frac{z-1}{2}$. The achievable secret key rate can then be expressed as~\cite{Devetak2005}:
\begin{align}
\label{DW_rate}
K_{\text{DW}} = \beta I_{AB} - \chi_{BE}.
\end{align}
 
We remark again that satellite communication links suffer from variation of the transmission efficiency due to atmospheric effects, variation of the link distance and fluctuation of the signal coupling in single-mode fiber. These fluctuations increase the variance of Bob’s measurement, leading to an additional term in the noise, known as fading noise $ \xi_{\text{fad}} = \frac{\mathrm{Var}(T)}{\mathbf{E}[T]^2} (V_A)$ \cite{dequal_feasibility_2021,Usenko2012}, where $\mathbf{E}[\cdot]$ is the expectation with respect to the fading process. As we have discussed earlier, to mitigate the detrimental effect of channel fading, it is possible to group the measurements according to the instantaneous channel transmission efficiency, which can be probed with a classical beacon. The number of groups is optimized with a trade-off between the fading reduction, which would require many groups, and finite-size effects, which requires few groups to have more symbols within each group.

One of the challenges of CV-QKD is sharing the local oscillator between Alice and Bob. This is required to have the same reference frame to encode and measure the phase information. In this work we exploit the so-called ``self-reference'' protocol \cite{Soh2015,Qi2015}, in which periodic intense signals are sent from Alice to Bob. From these pilots, Bob can recover the phase of Alice's local oscillator and correct his measurements accordingly. This process introduces two additional sources of noise, one related to the energy of the pilots, $E_{P}$ \cite{Soh2015}, and one related to a drift of the laser between the pilot and the quantum signal, $\Delta\nu_{P}$ \cite{Qi2015}. These two noise sources that depend on the pilot intensities and on the bandwidth of Alice and Bob's lasers, are considered in this analysis, and the reference values used for simulations are reported in the next section, together with other relevant parameters. Besides this noise contribution, we include other possible noise sources in the parameter $\xi$, the excess noise.

Finally, as done for the DV-QKD simulation we consider finite-size effects for CV as well. In particular, we consider the uncertainty of the parameter estimation due to the limited statistics ~\cite{Leverrier2010}. We can set a bound on the transmission coefficient $T$, \emph{i.e.}, the square root of the transmission efficiency $\tau$, and the  parameter $\sigma^2=1+\tau\xi$, which depends on the excess noise $\xi$, as:
\begin{align}
    T_{\text{min}}&\simeq\sqrt{\tau} - z_{\epsilon_{\mathrm{PE}}/2}\sqrt{\frac{1+\tau\xi}{m V_{\mathrm{A}}}} \\
    \sigma^2_{\text{max}}&\simeq1+\tau\xi+z_{\epsilon_{\mathrm{PE}}/2} \frac{(1+\tau\xi)\sqrt{2}}{\sqrt{m}},
\end{align}
where $m$ is the number of symbols used for parameter estimation and $z_{\epsilon_{\mathrm{PE}}/2}$ is a parameter related to the failing probability of the parameter estimation $\epsilon_{\mathrm{PE}}$. Here we consider $\epsilon_{\mathrm{PE}}=10^{-10}$, which gives $z_{\epsilon_{\mathrm{PE}}/2}= \sqrt{2} \; \textrm{erf}^{-1}(1-\epsilon_{\mathrm{PE}}) =6.5$, where $\textrm{erf}^{-1}$ is the inverse error function. 

\subsection{Simulation parameters}

We can now summarize all the simulation parameters used in our analysis. Part of the parameters describe features common to the two QKD systems, CV and DV. These values, reported in Table~\ref{tab:common_par}, refer to the satellite and ground station characteristics. 

For DV-QKD, the parameters used in the asymptotic simulation are summarized in Table~\ref{tab:DV_par} and correspond to the state-of-the-art free-space~\cite{Avesani2021} and satellite DV-QKD~\cite{liao_satellite--ground_2017}, at telecom wavelength, with $\eta_d$ corresponding to state-of-the-art superconducting nanowire detectors (SNSPD)~\cite{Avesani2021}, $f_{EC} = 1.16$~\cite{vasylyev_satellite-mediated_2019}, $e_d = 0.01$~\cite{vasylyev_satellite-mediated_2019,liao_satellite--ground_2017} and $q = 1$. We run the simulations for two values of the background noise $Y_0$, a typical one, corresponding to the estimated day ($1 \cdot 10^{-6}$) and night ($2 \cdot 10^{-7}$) background of a telescope placed in an urban area, and the other at $1.6 \cdot 10^{-4}$, corresponding to the limit case of a satellite reflecting sunlight onto the receiving telescope. The detailed calculation of these noise levels can be found in Appendix~\ref{SI:Background}.

For CV-QKD, the relevant parameters affecting the key rate both in the asymptotic regime and when considering finite-size effects are typical noise sources, such as electronic noise $v_{el}$ and phase recovery noise, as well as additional noise sources that cannot be determined \emph{a priori}, which we summarize here in the fixed excess noise parameter $\xi_{fix}$. A detailed calculation on how these parameters, summarized in Table~\ref{tab:CV_par}, enter the expression for the key rate can be found in \cite{dequal_feasibility_2021}.

It is worth underlying that since these values refer to different technologies needed for the realization of DV and CV-QKD terminals, a direct comparison of the performances of the two systems is not possible. 

\begin{table}[!ht]
    \centering
    \begin{tabular}{|l|c|c|}
    \hline
        \textbf{Common Parameter} & \textbf{Symbol} & \textbf{Reference value} \\ \hline
        Wavelength & $\lambda$ & 1550 nm\\
        Pointing error & $\theta_p$ & 1 $\mu$rad \\
        Divergence angle & $\theta_d$ & 10 $\mu$rad \\
        Fixed attenuation & $\eta_{opt}$ & 2.8 dB \\
         Zenith transmittance & $\tau_{\text{zen}}$ & 0.91 \\
        Transmission symbol rate & $f_{\mathrm{TX}}$ & 100 Msymbol/s\\
        Receiving telescope diameter  & $D_{TX}$ & 1.5 m\\
        \hline
    \end{tabular}
    \caption{Summary of the main simulation parameters used in our model, together with their reference baseline values.}
    \label{tab:common_par}
\end{table}

\begin{table}[!ht]
    \centering
    \begin{tabular}{|l|c|c|}
    \hline
        \textbf{DV Parameter} & \textbf{Symbol} & \textbf{Reference value} \\ \hline
        Background noise & $Y_0$ &  $1 \cdot 10^{-6}$/$2 \cdot 10^{-7}$ \\
        & & $1.6 \cdot 10^{-4}$ \\
        Detection efficiency & $\eta_d$ & $0.85$ \\
        Prob. of erroneous detection & $e_d$ & $0.01$ \\
        Asym. prob. of Z basis & $q$ & $1$ \\
        Correctness parameter & $\varepsilon_{corr}$ & $10^{-10}$ \\
        Security parameter & $\varepsilon_{sec}$ & $10^{-10}$ \\
        \hline
    \end{tabular}
    \caption{Summary of the main parameters used for DV-QKD simulation, together with their reference values.}
    \label{tab:DV_par}
\end{table}

\begin{table}[!ht]
    \centering
    \begin{tabular}{|l|c|c|}
    \hline
        \textbf{CV Parameter} & \textbf{Symbol} & \textbf{Reference value} \\ \hline
        Electronic noise & $v_{\mathrm{el}}$ & 10\% S.N.U.\\
        Detection efficiency & $\eta $ & $0.4 $\\
        Fixed excess noise & $\xi_{\mathrm{fix}}$ & 1-5\% S.N.U.\\
        Pilot energy &$E_{P}$ & 10 pJ\\
        Pilot bandwidth &$\Delta\nu_{P}$ & 10 kHz\\
        Reconciliation efficiency & $\beta$ & 0.95 \\
        \hline
    \end{tabular}
    \caption{Summary of the main parameters used for CV-QKD simulation, together with their reference values. }
    \label{tab:CV_par}
\end{table}

\section{Sensitivity to critical parameters}

At this point, we have laid the ground for assessing the feasibility of satellite-to-ground QKD for a general scenario, both in CV and DV. We complete this study by assessing the sensitivity of the QKD performance to critical parameters of the scenario, namely the turbulence strength, the satellite altitude, the ground receiving telescope diameter and the optimization of transmitted bits grouping.

\subsection{Sensitivity to turbulence strength}

\begin{figure}
\centering
  \includegraphics[width=8cm]{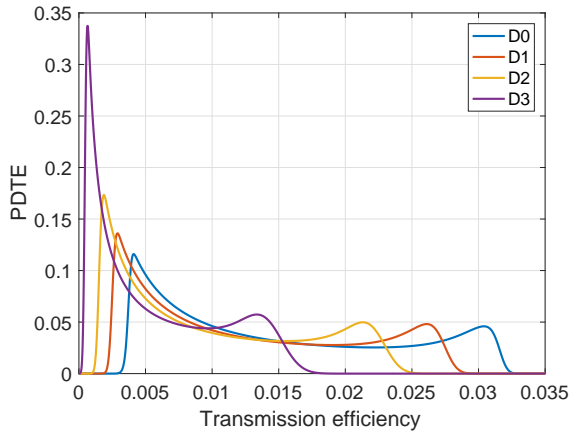}
\caption{Probability distributions of the transmission efficiency for several daylight turbulence profiles for a 500 km orbit with an AO correcting 15 orders. }
\label{PDTE_vs_turbulence}
\end{figure}

We study first the sensitivity of the key rate performance to the turbulence strength and the mitigation effect of several AO systems, with a number of corrected Zernike radial orders ranging from 5 to 20. Following the procedure detailed in Section~\ref{PDTE_calc}, we calculated the PDTE for day and night, considering eight total turbulence profiles, ranging from mild to severe turbulence including our median reference cases. Their integrated parameters for a $90^\circ$ elevation are detailed in Tables~\ref{int_parametersD} and \ref{int_parametersN}. We limited our analysis to a satellite altitude of 500 km. The resulting PDTEs are shown in Fig.~\ref{PDTE_vs_turbulence} for the four daytime turbulence profiles and an AO system correcting 15 orders.

 \begin{table}[h]
        \centering
         \begin{ruledtabular}
        \begin{tabular}{ccccc}
    \hline
    & $D_0$ &$D_1$&$D_2$&$D_3$\\
     \hline
      $r_0$ (cm) & 24.8 &15.0&10.6&6.9\\
     \hline
     $\tau_0$ (ms) &2.4 &1.8&1.3&0.9\\
     \hline
      $\theta_0$ ($\mu$rad) &45.8 &34.5&25.8&18.1\\
     \hline
      $\sigma_\chi ^2$ &0.005 &0.01&0.01&0.02\\
     \hline
    \end{tabular}
    \end{ruledtabular}
    \caption{Integrated parameters calculated at at 90$^\circ$ for the different daytime turbulence profiles, $D_1$ being the reference profile previously used in the study. }
    \label{int_parametersD}
\end{table}

 \begin{table}[h]
        \centering
         \begin{ruledtabular}
        \begin{tabular}{ccccc}
    \hline
    & $N_0$ &$N_1$&$N_2$&$N_3$\\
     \hline
      $r_0$ (cm) &68.6 &50.4&37.8&22.9\\%9.9 & 16.4 & 7.0 &4.5\\
     \hline
      $\tau_0$ (ms) & 2.4&1.8&1.3&0.9\\
      \hline
      $\theta_0$ ($\mu$rad) &45.9 &34.4&25.9&18.1\\%& 11.3  &15.2 &8.5&6.0\\
     \hline
      $\sigma_\chi ^2$ &0.0045 &0.008&0.01&0.02\\%& 0.03  & 0.02 & 0.05& 0.08\\
     \hline
    \end{tabular}
    \end{ruledtabular}
    \caption{Integrated parameters calculated at at 90$^\circ$ for the different nighttime turbulence profiles, $N_1$ being the reference profile previously used in the study.  }
    \label{int_parametersN}
\end{table}

\begin{figure}[h!]
\centering
  \includegraphics[width=8cm]{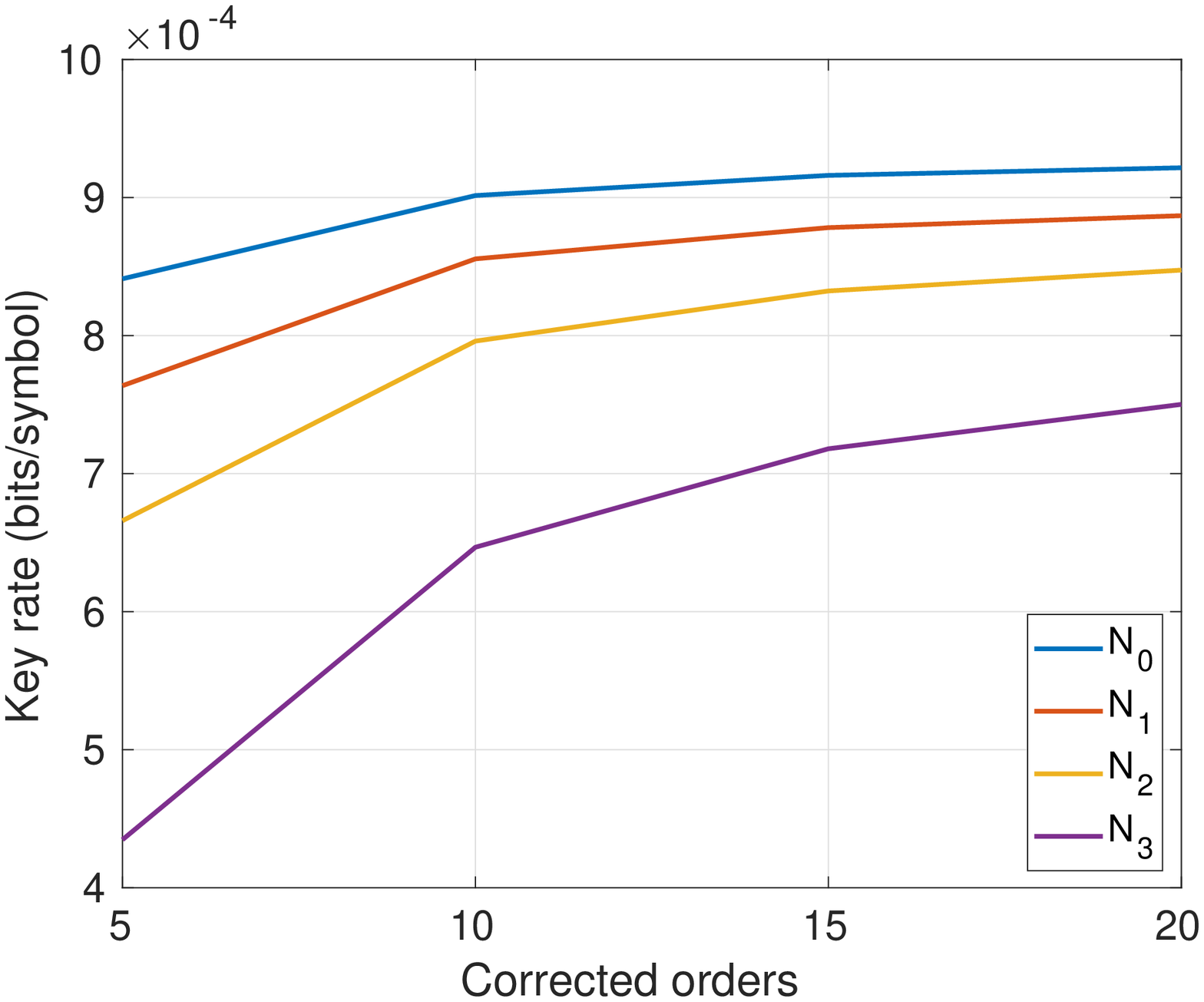}
  \includegraphics[width=8cm]{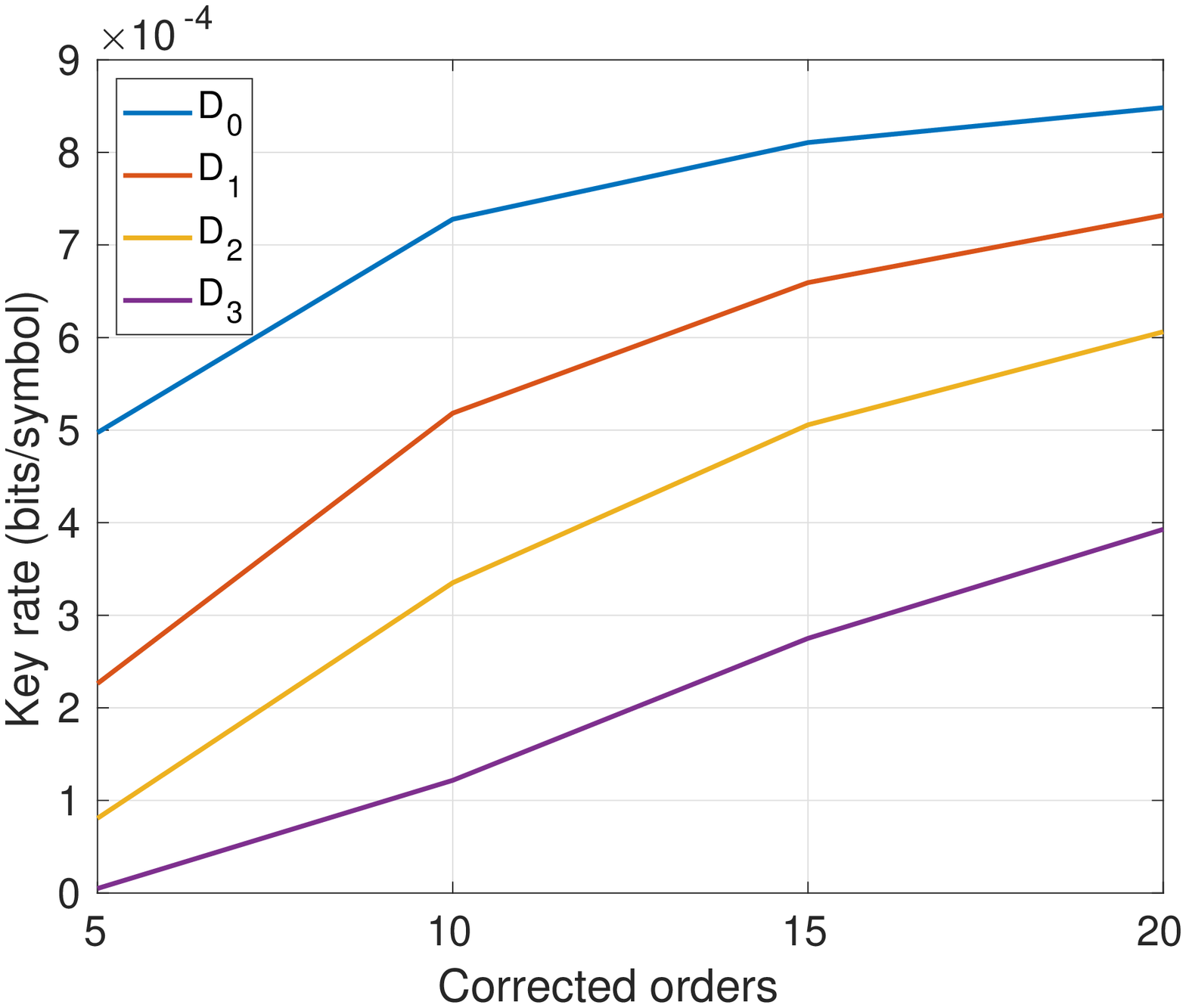}
\caption{DV-QKD secret key rate, including finite-size effects, for a 500 km orbit and several  turbulence regimes at night (top) and day (bottom) time, as a function of the AO correction orders.}
\label{turbulence_DV}
\end{figure}

\begin{figure}[h!]
\centering
  \includegraphics[width=8cm]{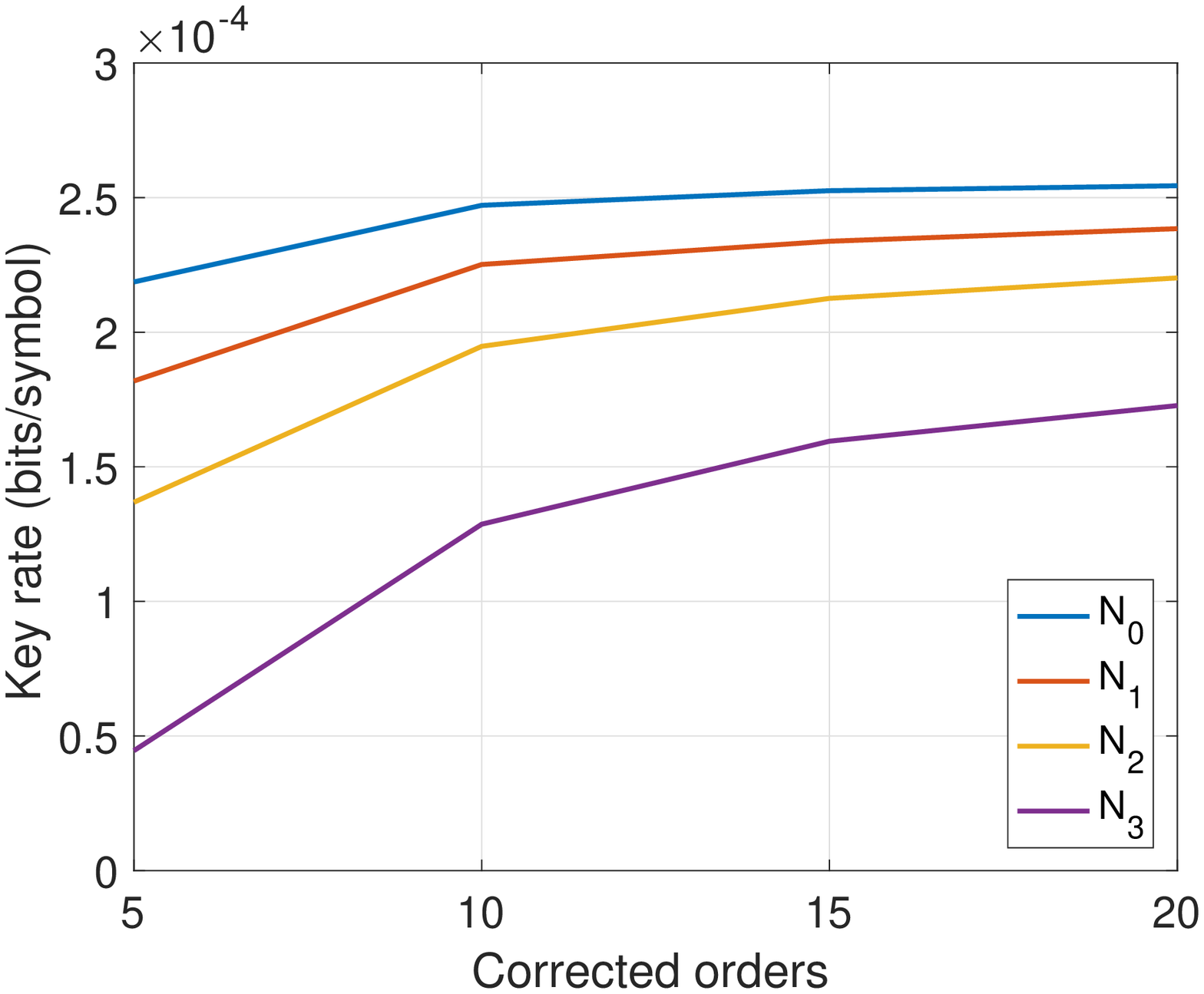}
\includegraphics[width=8cm]{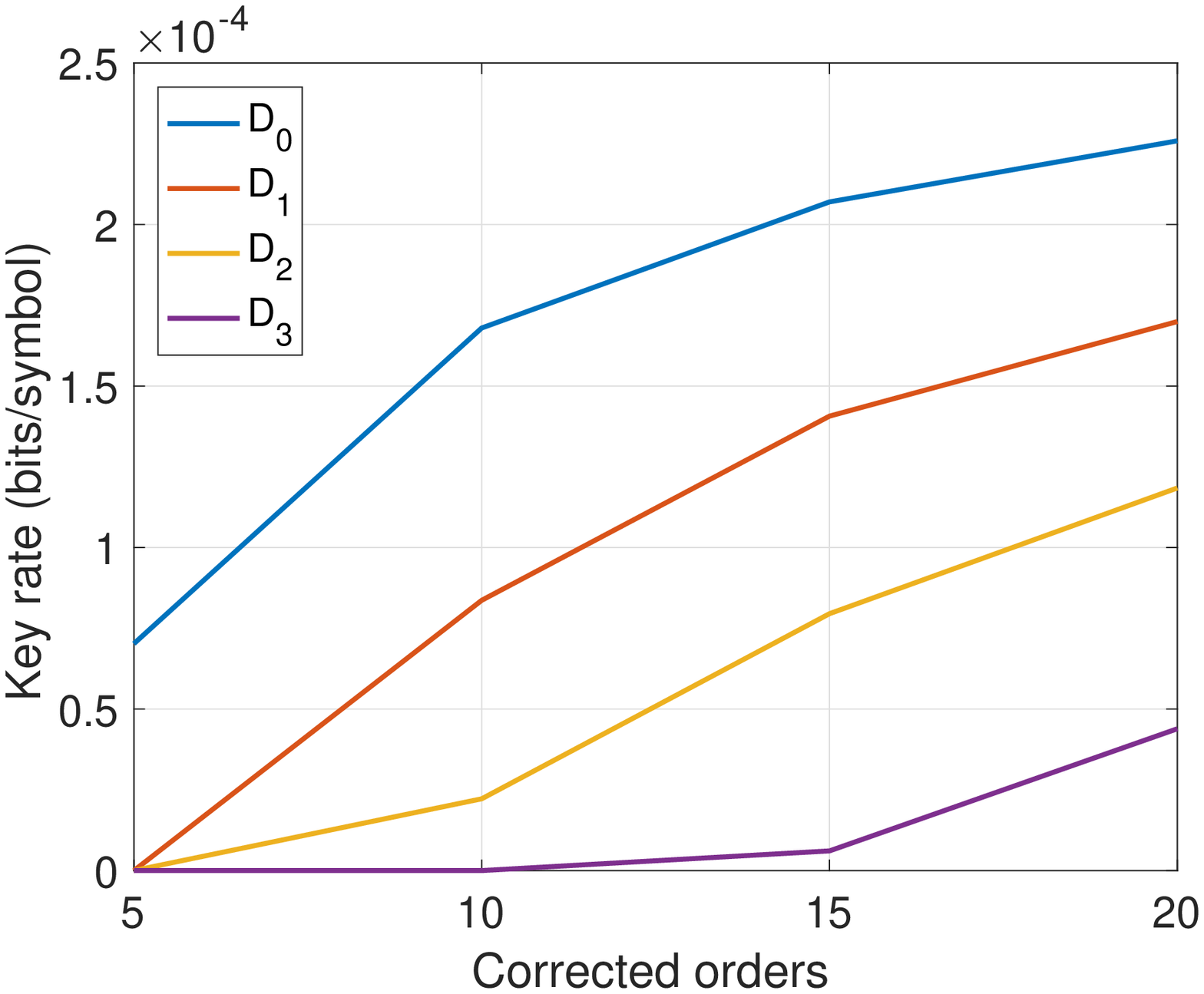}
\caption{CV-QKD secret key rate, including finite-size effects, for a 500 km orbit and several  turbulence regimes at night (top) and day (bottom) time, as a function of the AO correction orders.}
\label{turbulence_CV}
\end{figure}

On the basis of the PDTE obtained, we estimated the key rate for both DV and CV-QKD, including finite-size effects. As shown in Figs. \ref{turbulence_DV}  and \ref{turbulence_CV} (top), for nighttime turbulence profiles the key rate improves when increasing the number of corrected orders from 5 to 10. For higher orders, however, the improvement is marginal, in particular for milder turbulence. This trend can be explained by the fact that these turbulence profiles introduce a limited amount of wavefront distortion and lower order AO can already compensate for it, reaching an almost optimal coupling efficiency. 

On the contrary, for day profiles the key rate improves steadily with the number of corrected orders up to 20, as shown in Figs. \ref{turbulence_DV} and \ref{turbulence_CV} (bottom). This is particularly true for more severe turbulence regimes, as expected.

It is interesting to note that, while the optimal AO system strongly depends on the atmospheric condition, a system capable of correcting up to 15 orders could reach an almost optimal fiber coupling in most nighttime conditions and would have a slightly sub-optimal coupling in daylight for mild to medium turbulence. Since such an AO system is within the reach of the technological development \cite{kernec_h2020_2021}, we will consider this scenario as a reference in the remaining analysis.

\subsection{Sensitivity to satellite altitude}

\begin{figure}[h!]
\centering
%\begin{subfigure}%{.5\textwidth}
 % \centering
  \includegraphics[width=8cm]{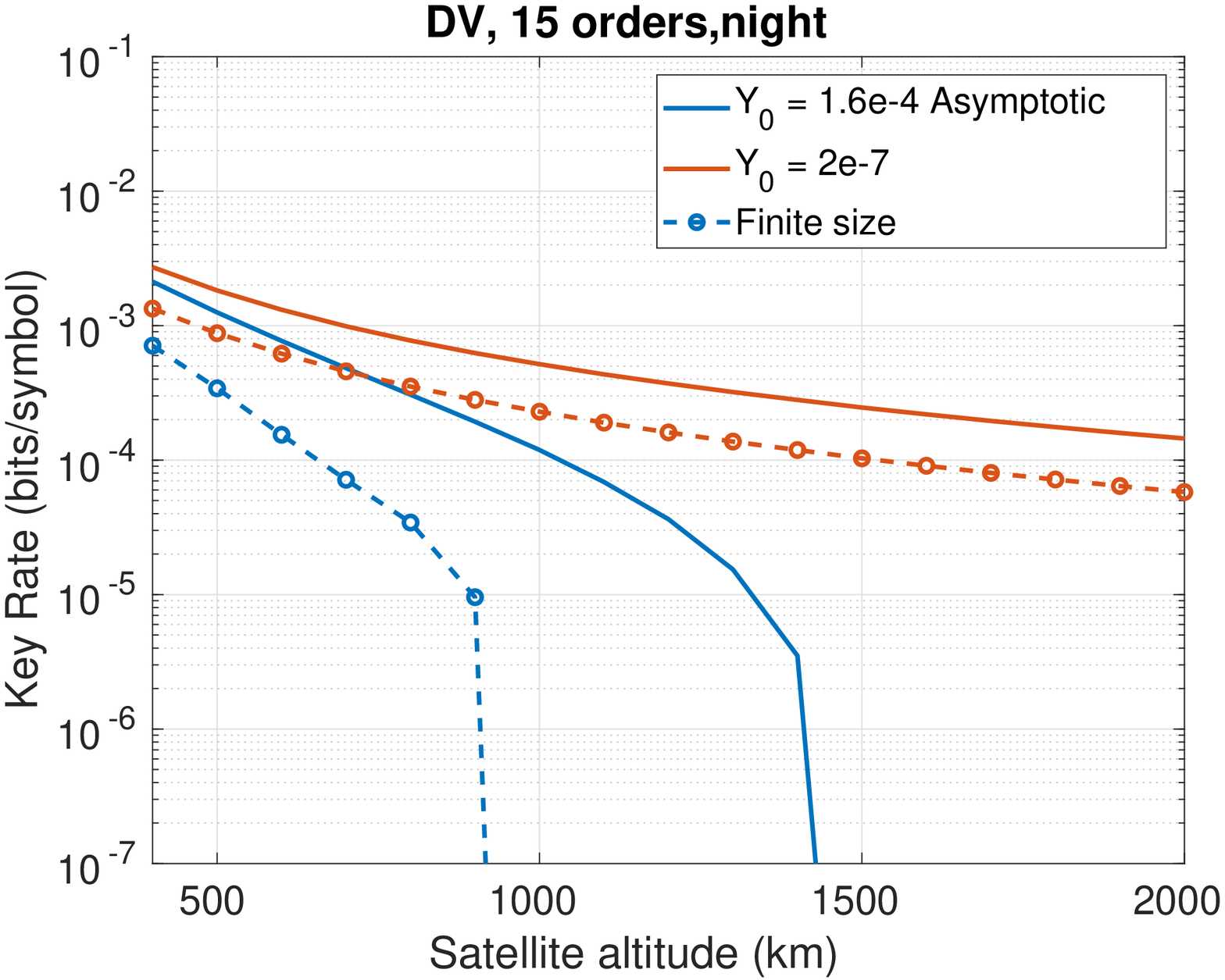}
%\end{subfigure}%
%\begin{subfigure}%{.5\textwidth}
 % \centering
  \includegraphics[width=8cm]{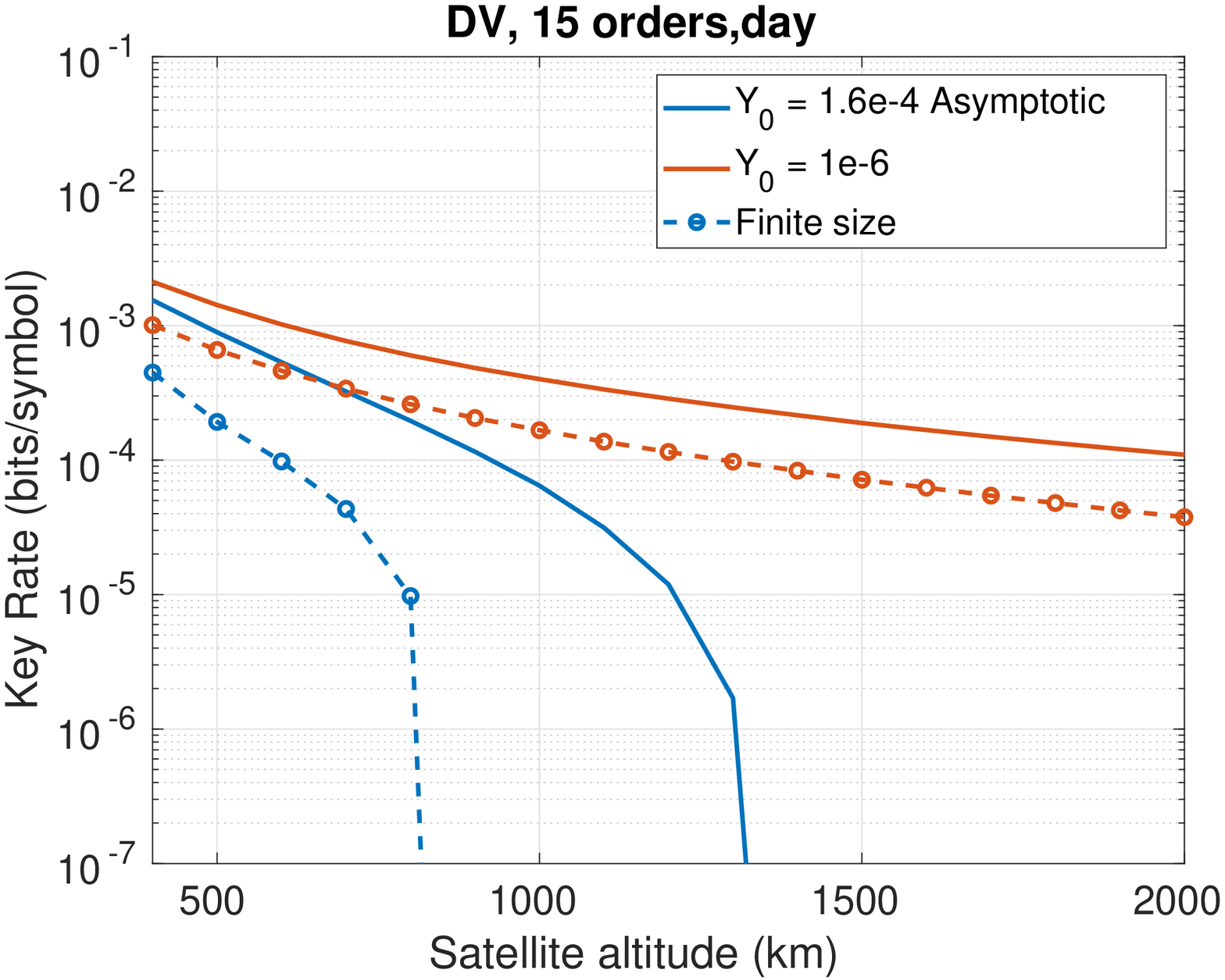}
%\end{subfigure}
\caption{DV-QKD secret key rate in the asymptotic regime (solid lines) and including finite-size effects at a 100 MHz rate (dashed lines), as a function of satellite altitude. Simulations performed using night (top) and day (bottom) N1 and D1 turbulence profiles. }
\label{fig_avg_Teff2}
\end{figure}

\begin{figure}[h!]
\centering
%\begin{subfigure}%{.5\textwidth}
 % \centering
  \includegraphics[width=8cm]{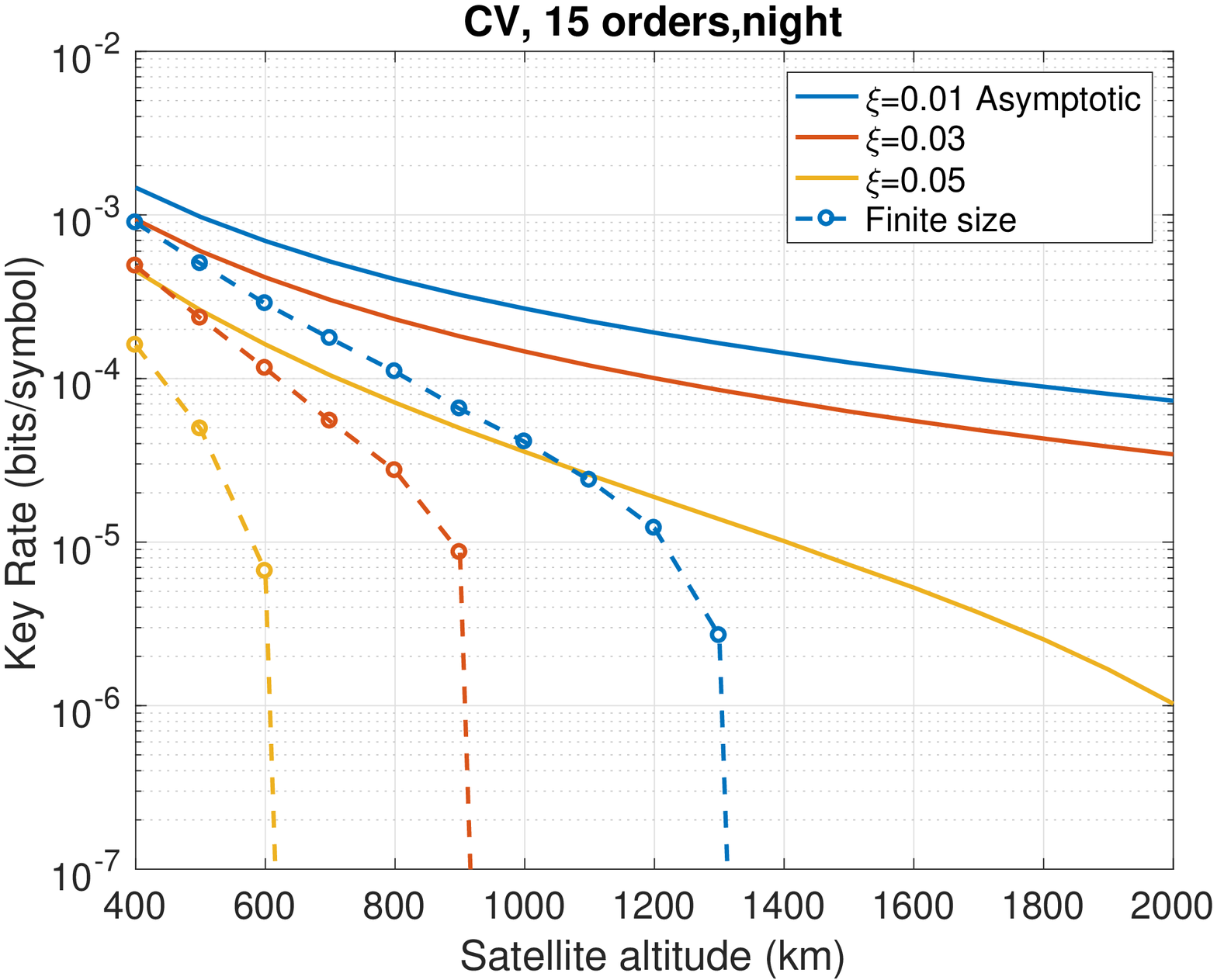}
%\end{subfigure}%
%\begin{subfigure}%{.5\textwidth}
 % \centering
  \includegraphics[width=8cm]{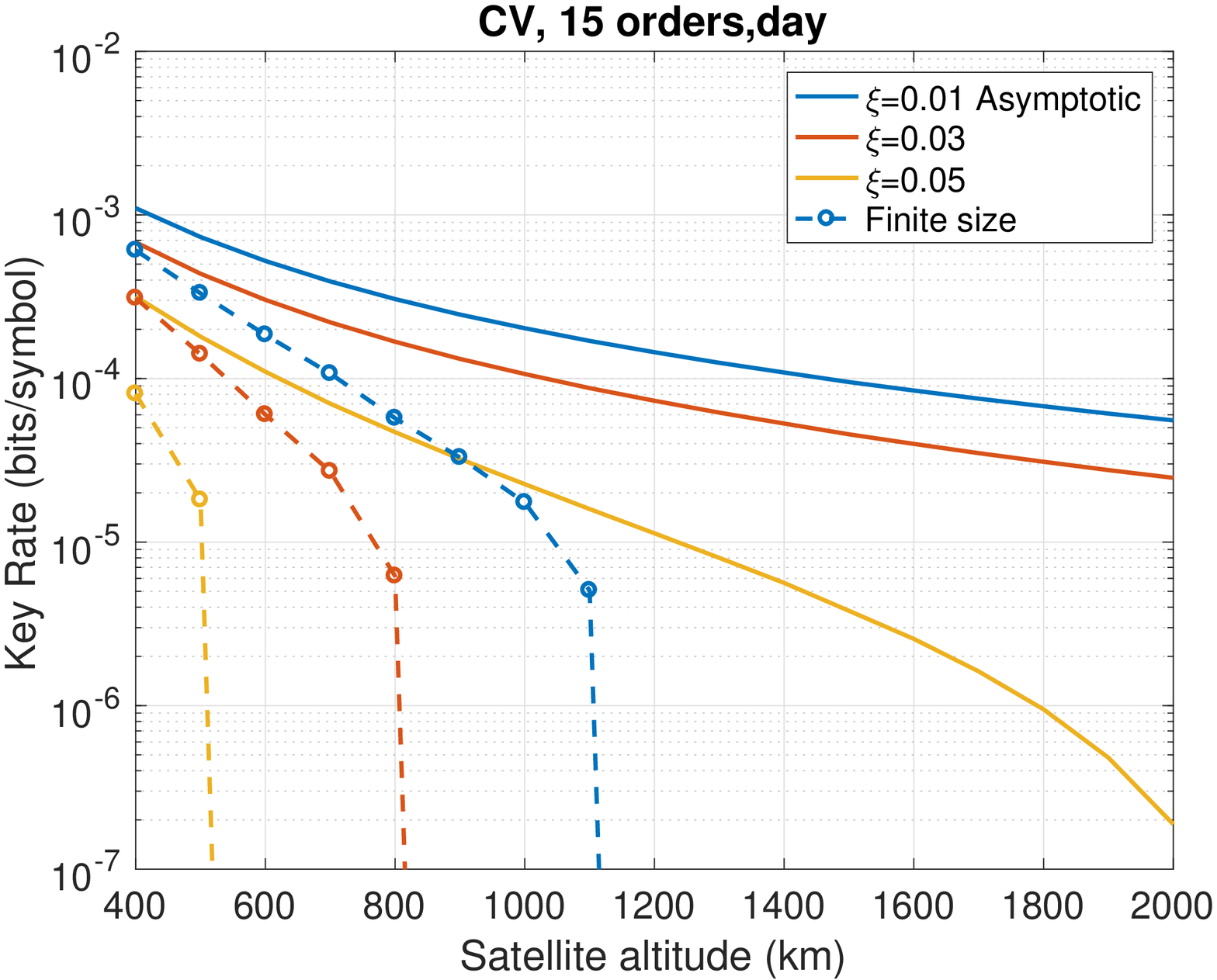}
%\end{subfigure}
\caption{CV-QKD secret key rate in the asymptotic regime (solid lines) and including finite-size effects (dashed lines), as a function of satellite altitude. Simulations performed using night (top) and day (bottom) time N1 and D1 turbulence profiles.}
\label{fig_avg_Teff}
\end{figure}

We now assess the sensitivity of the secret key rate to the satellite altitude. Figures \ref{fig_avg_Teff2} and \ref{fig_avg_Teff} represent the key rate for DV and CV-QKD, respectively, assisted by an AO system correcting 15 radial orders, and operating during at day and night time, for satellite altitudes ranging from 400 km to 2000 km. Both the asymptotic regime and including finite-size effects are shown.
For CV, we consider excess noise values of 1\%, 3\% and 5\%, and for DV, a typical 0-photon yield of $1 \cdot 10^{-6}$ and $2 \cdot 10^{-7}$ for, respectively at day time and night time, as well as a pessimistic case of $1.6 \cdot 10^{-4}$. As expected, the key rate performance degrades as the satellite altitude increases. 

In the case of DV, positive secret key rate can be obtained from all LEO altitudes in the case of lower background noise, both for the asymptotic regime and when considering finite-size effects. In case of more severe background noise, the key exchange is limited to lower orbits, both for finite-size and asymptotic regime. It is worth noting, however, that the reach of DV-QKD can be extended to a certain amount by increasing the number of symbols considered.

In the case of CV, the asymptotic key rate shows a similar behaviour for day and night, giving a non null key rate for all LEO altitudes. Instead, when considering finite-size effects the two scenarios clearly differ. In particular, for daylight turbulence profiles no key is possible for an excess noise of $\xi=0.05$, and the reach for lower noise is reduced. It is worth noticing that in the CV case, the finite-size effects have a greater impact on the feasibility of satellite QKD, in particular for higher orbits and noise levels. As for DV-QKD, this could be mitigated by increasing the repetition rate of the system or by merging together several satellite passes.

\subsection{Analysis for an 80 cm telescope}

\begin{figure}%[hbt!]
\centering
%\begin{subfigure}%{.5\textwidth}
 % \centering
  \includegraphics[width=8cm]{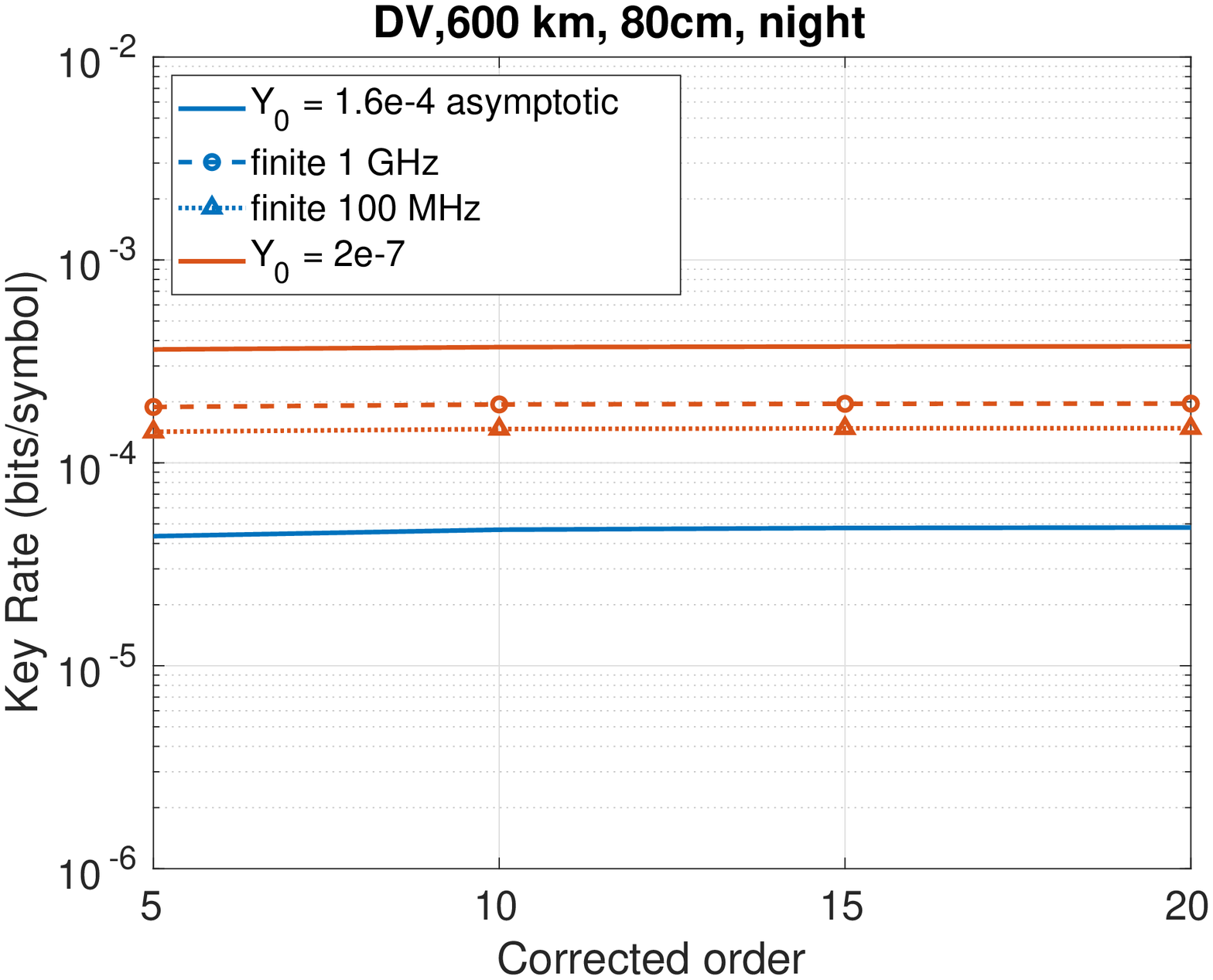}
%\end{subfigure}%
%\begin{subfigure}%{.5\textwidth}
 % \centering
  \includegraphics[width=8cm]{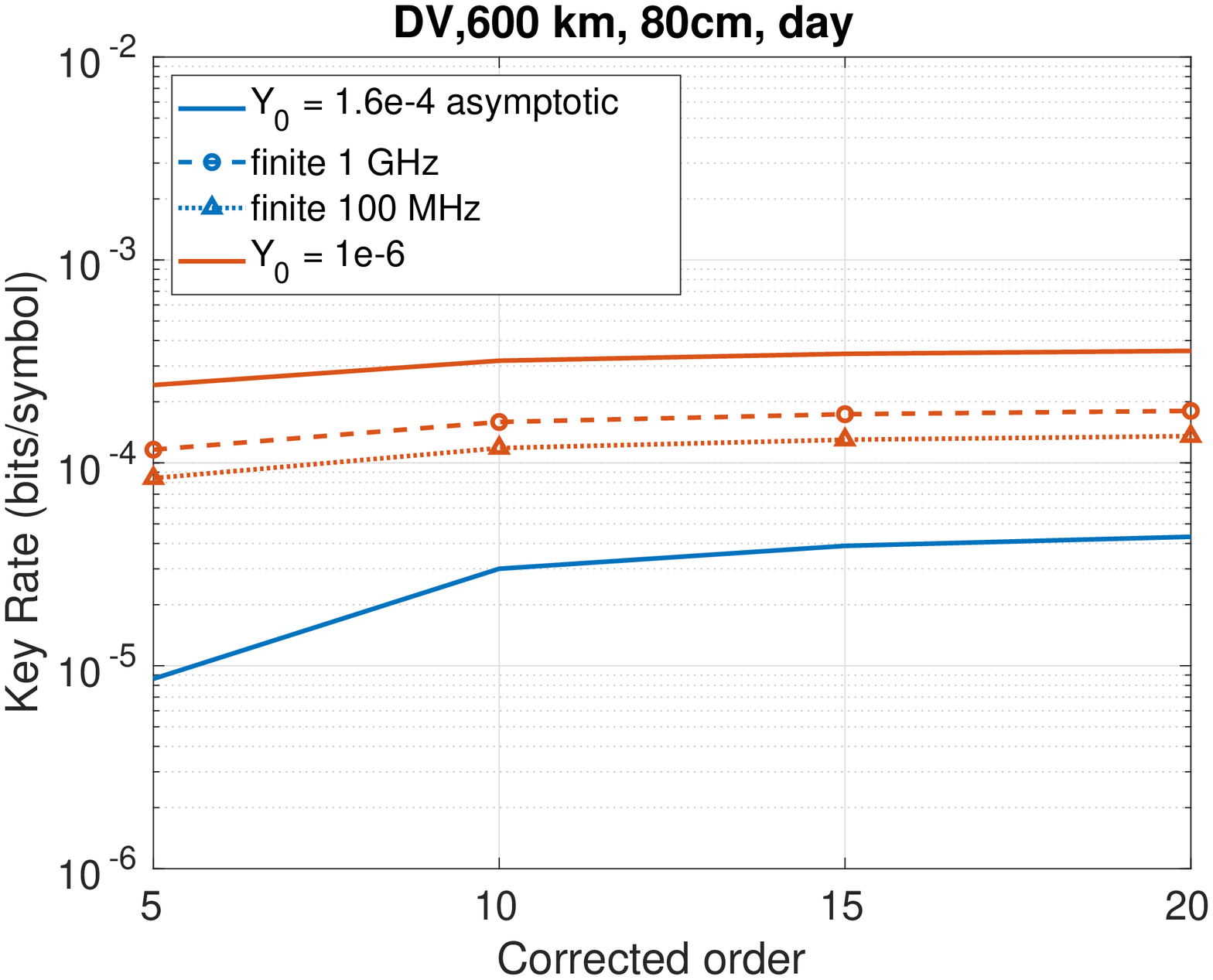}
%\end{subfigure}
\caption{DV-QKD secret key rate in the asymptotic regime (solid lines) and including finite-size effects at 1 GHz (dotted lines) and 100 MHz (dashed lines) for a 80 cm receiving telescope, as a function of the AO corrected orders. Simulations performed using night N1 (top) and day D1 (bottom) turbulence profiles.}
\label{fig_avg_Teff4}
\end{figure}

\begin{figure}%[hbt!]
\centering
%\begin{subfigure}%{.5\textwidth}
 % \centering
  \includegraphics[width=8cm]{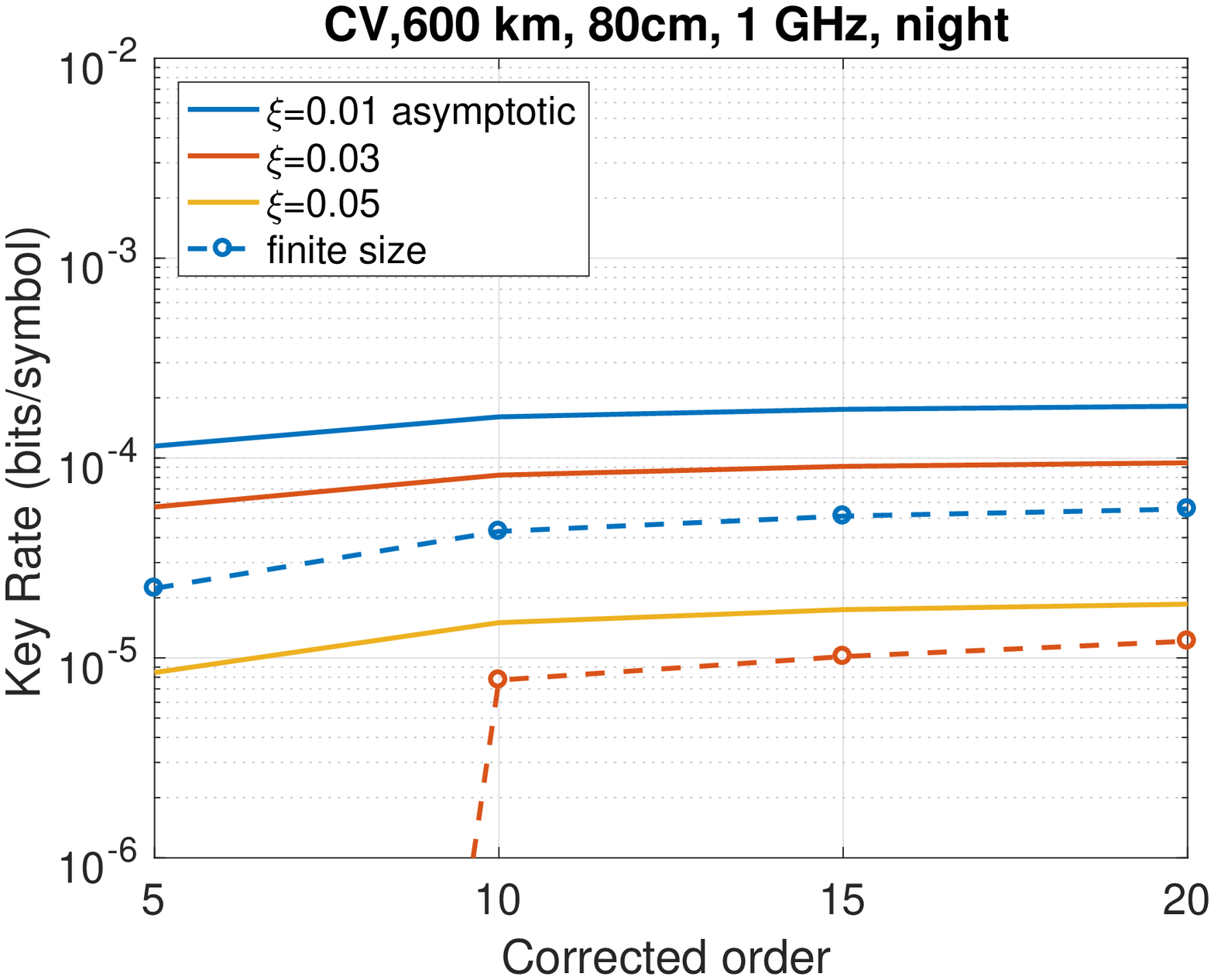}
%\end{subfigure}%
%\begin{subfigure}%{.5\textwidth}
 % \centering
  \includegraphics[width=8cm]{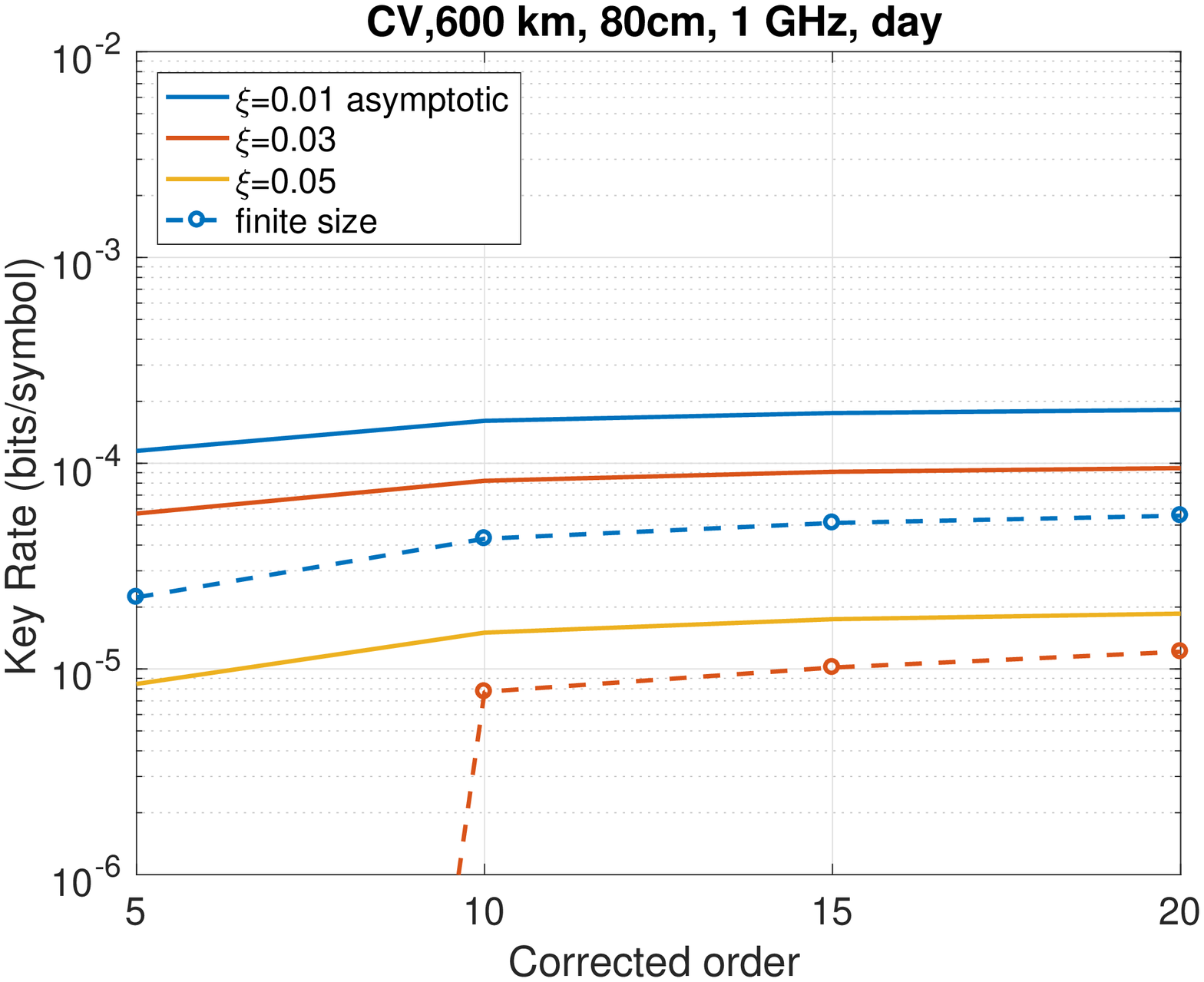}
%\end{subfigure}
\caption{CV-QKD secret key rate in the asymptotic regime (solid lines) and including finite-size effects (dashed lines) for a 80 cm receiving telescope and 1 GHz of symbol rate, as a function of the AO corrected orders. Simulations performed using night N1 (top) and day D1 (bottom) turbulence profiles.  }
\label{fig_avg_Teff3}
\end{figure}

We continue our analysis by estimating the secret key rate assuming a ground telescope diameter of 80 cm instead of 1.5 m as was considered before. The satellite altitude is fixed at 600 km. 
Figures \ref{fig_avg_Teff4} and \ref{fig_avg_Teff3} show the results for DV and CV-QKD, respectively. 

For DV, in the case of stronger noise level, key rate can only be achieved in the asymptotic regime. Instead, for lower background noise the key rate can also be achieved in the finite-size regime, for both 100 MHz and 1 GHz sources. The achieved key rates are slightly better in night operation, and in this case are completely independent of the AO configuration. During day operation, AO correction provides a gain of a factor 2 on the asymptotic key rate assuming a 0-photon yield of $1.6e^{-4}$. It appears that complex AO systems are not required for smaller telescope sizes in DV-QKD.

For CV, no secret key can be transmitted at a 100 MHz rate for this telescope size, whatever the AO configuration, and increasing the transmission rate to 1 GHz is thus necessary to enable key distribution, reflecting once again the importance of finite-size effects for CV. For night time, the secret key rate is almost constant with respect to the AO configuration, indicating that for an 80 cm telescope with weaker turbulence conditions an advanced AO system is unnecessary. For daylight turbulence profiles in the asymptotic regime, correcting 10 radial orders provides a gain of a factor $<2$ on the key rate compared to a 5 radial order correction, and this performance is only slightly improved by correcting 15 or 20 radial orders, again indicating that advanced AO systems are not essential for 80 cm telescopes. When accounting for finite-size effects, CV-QKD is not feasible in the worst case of 5\% excess noise, and is only possible in the case of 3\% excess noise when correcting at least 10 radial orders.

%For both DV and CV. Might be interesting to see which divergence would allow us to make key with a 80 cm telescope.

\subsection{Optimization of grouping}

As explained in Section~\ref{key_rate_section}, to mitigate the detrimental effect of fluctuations of the transmission efficiency we divided the overall PDTE in a certain amount of groups, for both DV and CV-QKD analysis. We then treat each group independently, performing a complete key rate estimation for each group. Here, we discuss the optimal number of groups required in DV and CV.

For DV-QKD, the optimal number of groups is always smaller than 5 in all conditions, with the lower orbits requiring less subdivisions, in particular for the low noise configuration. This can be explained by the fact that for DV the only effect of channel fluctuations is a variation of the signal to noise ratio (SNR). Therefore, only when the SNR becomes too small (\emph{i.e.}, for higher orbits and higher level of dark counts) it is favorable to subdivide the PDTE to remove the low SNR sections. In the remaining cases, \emph{i.e.}, those with high SNR, it is more favorable to perform few (or no) subdivisions, as in this case the number of symbols used for parameter estimation increases, thus reducing finite-size effects. 

For CV-QKD however, as the signal is encoded also in light amplitude, channel fluctuations directly introduce a noise source. The channel variation is higher for lower orbits, due to a larger distance excursion between the zenith and $20^\circ$ elevation. In this case the optimal subdivision can reach 10 groups for 400 km, with this number decreasing with satellite altitude. However, for higher orbits while the distance excursion is smaller, the average attenuation is higher, making finite-size effects more critical. 

\section{Conclusion}

The realization of an intercontinental network of QKD nodes through satellites requires to address several technological challenges, in order to improve the communication link performance as well as their reliability. Among these challenges, the mitigation of detrimental effects of atmospheric turbulence on signal propagation represents a crucial challenge, in particular when single-mode fiber coupling is required, e.g., for compatibility with CV and DV-QKD detectors or as a strong spatial mode filtering to reduce atmospheric background.

In this work we address this topic by investigating the influence of an AO system in satellite QKD key rate generation, both with DV and CV protocols and considering finite-size effects. We started by modeling the transmission channel, taking into account geometrical losses, atmospheric attenuation, turbulence-induced wavefront perturbation and the compensation capabilities of several AO apparatuses. By considering all these effects, we reconstructed the probability distribution of the transmission efficiency for a complete satellite orbit as a function of several turbulence regimes and AO characteristics. We extended the analysis to consider satellite altitudes spanning all the LEO region, from 400 km to 2000 km. 

The results of our simulations show the fundamental improvement obtained by using an advanced AO system for fiber coupling in all communication scenarios, from mild to strong turbulence regimes in both DV and CV encoding, when a large 1.5 m aperture is used as a receiving station. On the other hand, the use of smaller receiver, with a diameter of 80 cm, shows a minor improvement in using advanced AO systems, in particular for nighttime conditions, which are characterized by lower level of turbulence. The simple case of using a tip tilt AO system is discussed in Appendix~\ref{SI:TipTilt}. 

The limits of feasibility for both DV and CV-QKD from satellite have been assessed under assumptions compatibles with state of the art technology for what concerns spacecraft and ground segment terminals. For DV, a clear dependence on background noise has been identified which limits the feasibility of QKD to lower orbits. For CV, a range of excess noise compatible with satellite QKD has been identified. The limitations related to finite-size effects are more severe in this case compared to. DV-QKD, and limit the feasibility of CV-QKD to satellite altitudes lower than 1200 km, even in the best configuration considered.

\begin{acknowledgments}
The authors wish to thank J. Osborn for kindly providing access to $C_n^2$ experimental data basis \cite{osborn_optical_2018}. We acknowledge support from the European Commission’s Horizon 2020 Research and Innovation Program under Grant Agreement no. 820466 (CiViQ).
\end{acknowledgments}

\bibliographystyle{apsrev4-1}
\bibliography{main}

\appendix

\section{Supplementary material}

%\section{\label{sec:level1}Supplementary information}

\subsection{Tip Tilt analysis}
\label{SI:TipTilt}

Here we wish to study the effect of a simpler AO system, based only on tip-tilt correction, on the feasibility of satellite QKD. While the complexity of such an apparatus is greatly reduced with respect to higher orders AO, it offers limited wavefront correction and consequently single-mode fiber injection efficiency. 
In this analysis, we consider two standard turbulence profiles for nigh time and daylight, namely $N_1$ and $D_1$. 

\begin{figure}[hbt!]
\centering
  \includegraphics[width=8cm]{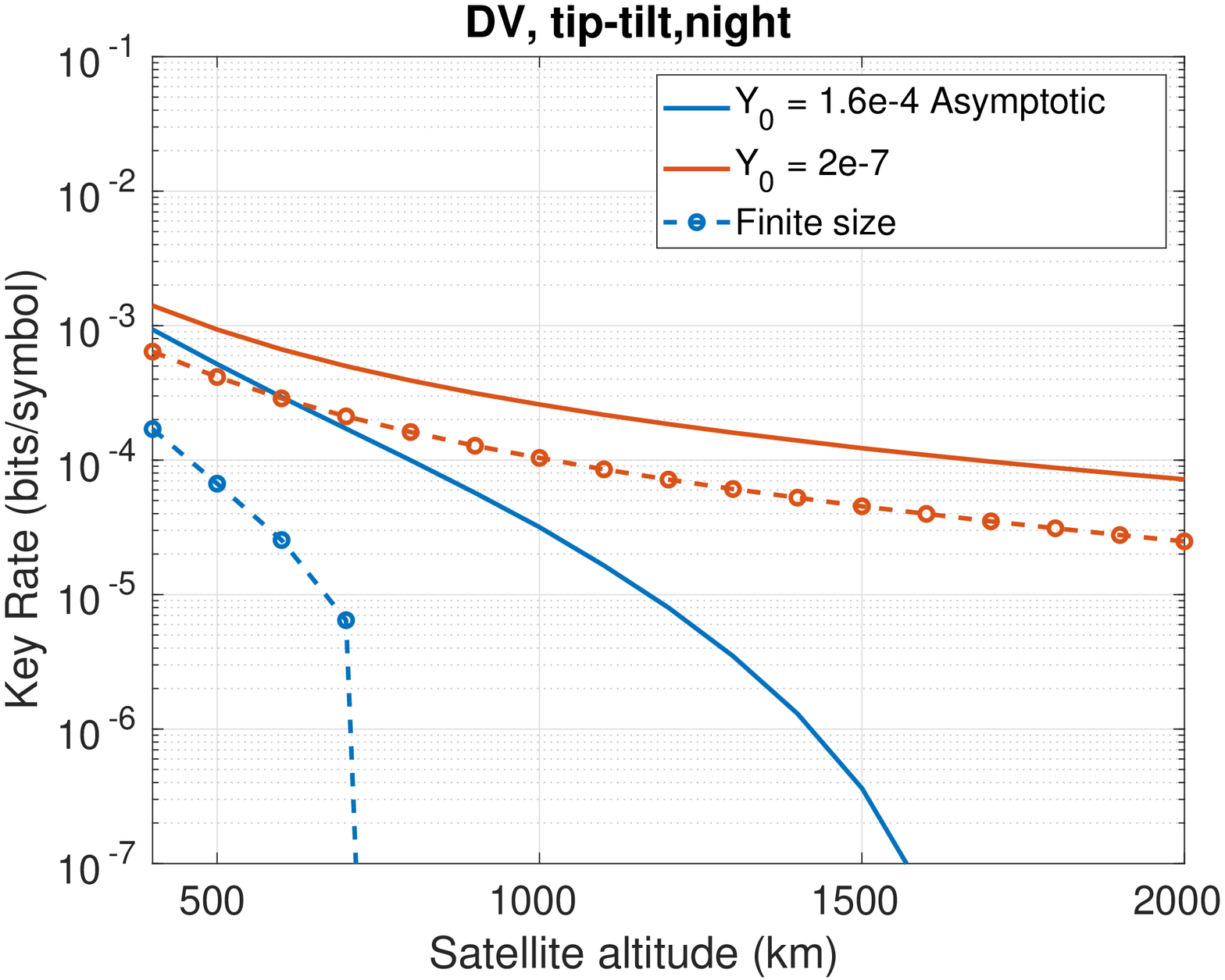}
  \includegraphics[width=8cm]{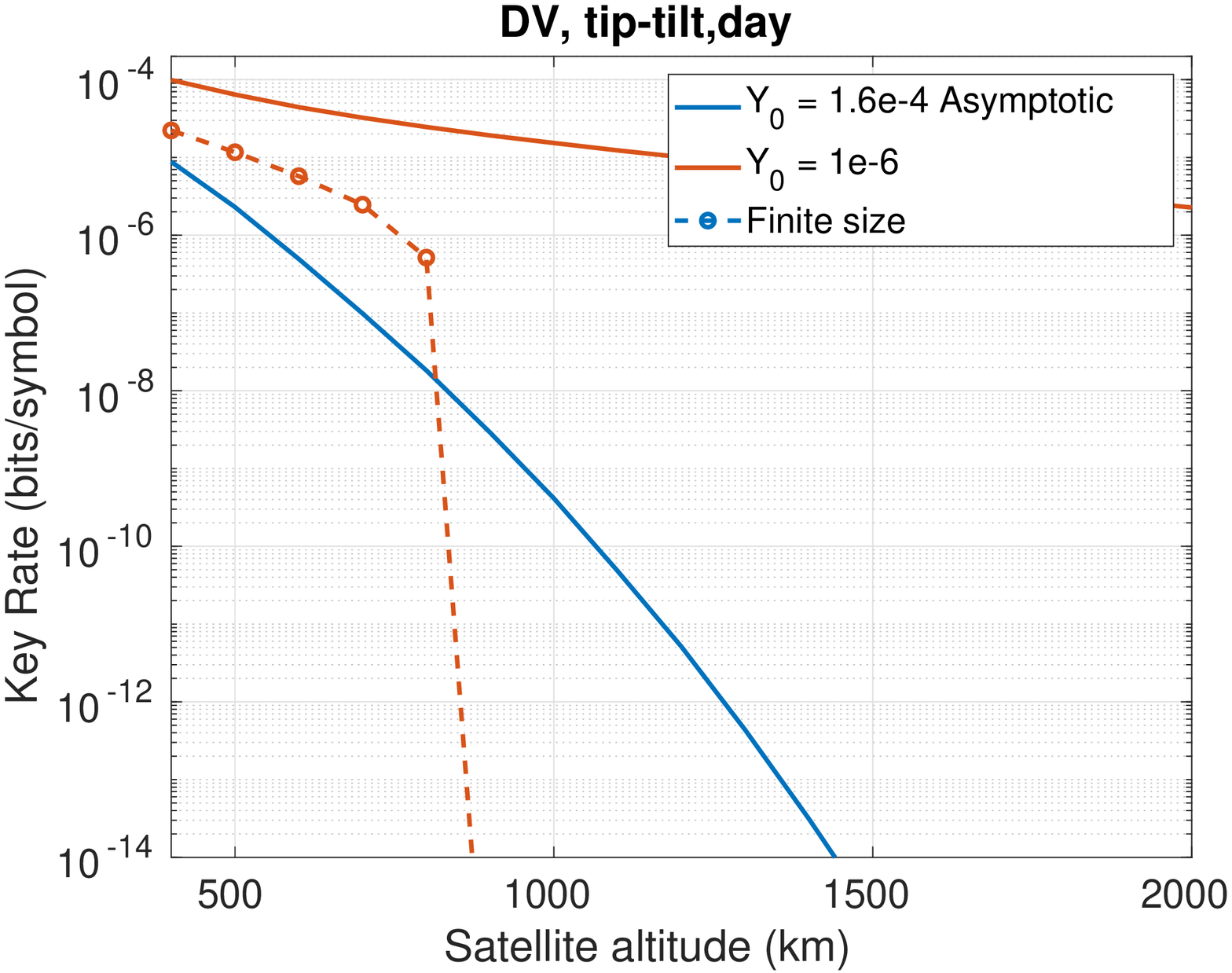}
\caption{DV-QKD secret key rate as a function of satellite altitude with AO correcting only the first order (\emph{i.e.}, with tip-tilt correction), for standard  turbulence regimes at night (top) and day (bottom) time.}
\label{tiptilt_DV}
\end{figure}

For DV-QKD, QKD is feasible at night time even with only 1 correction order for AO, with just a slight reduction in the secret key rate as shown in Fig.~\ref{tiptilt_DV}. For daylight, high order correction for adaptive optics is necessary, since with 1 order correction it is possible to obtain a key, when finite-size effects are considered, only for satellite orbits lower than 800 km at the typical background noise level, while no key can be obtained for the pessimistic noise regime.

\begin{figure}[hbt!]
\centering
  \includegraphics[width=8cm]{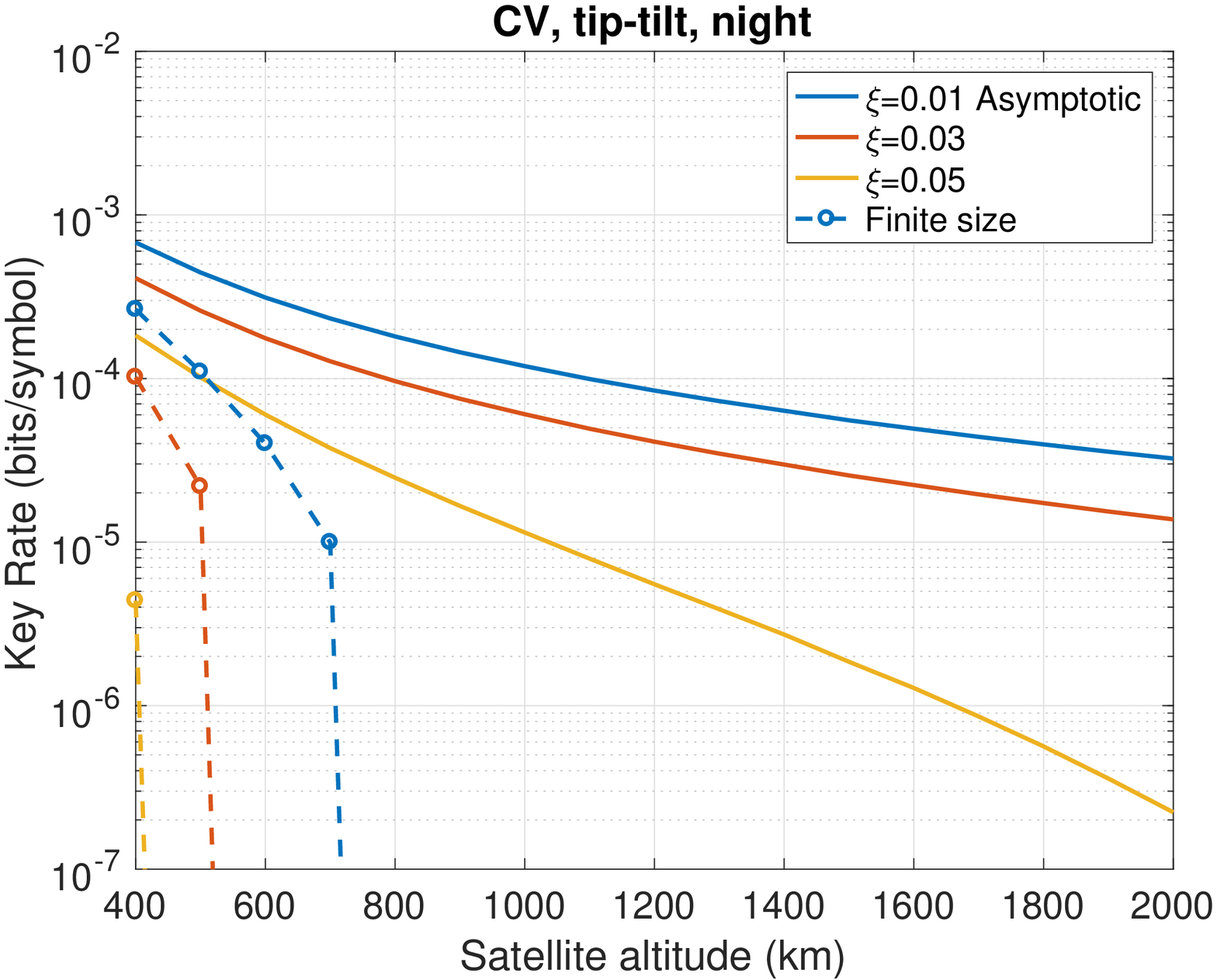}
  \includegraphics[width=8cm]{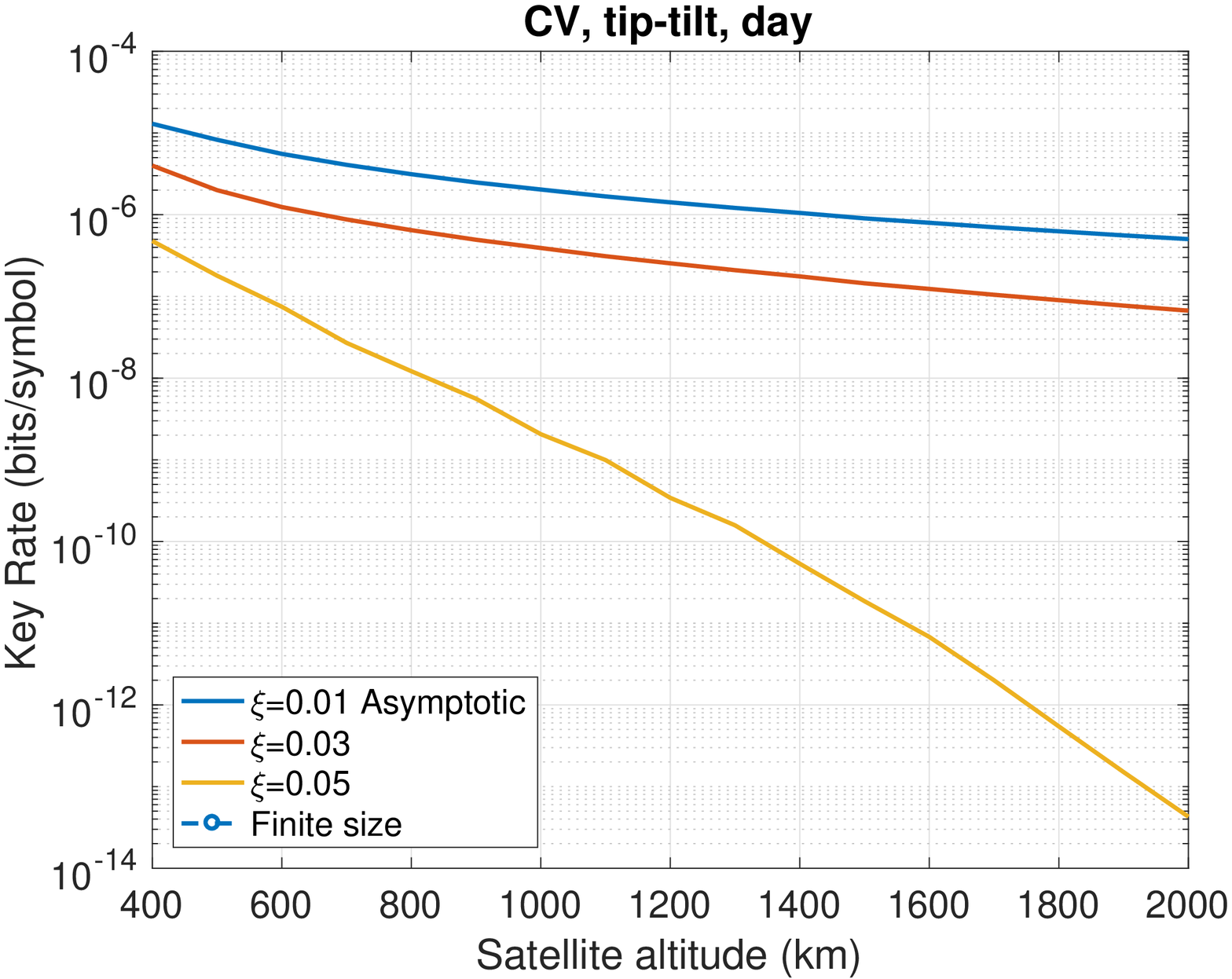}
\caption{CV-QKD secret key rate as a function of satellite altitude with AO correcting only the first order (\emph{i.e.}, with tip-tilt correction), for standard  turbulence regimes at night (top) and day (bottom) time.}
\label{tiptilt_CV}
\end{figure}

For CV, as shown in Fig.~\ref{tiptilt_CV}, the use of 1 correction order for AO drastically reduces the feasibility of QKD, which is not possible for any orbit higher than 700 km for night time for any noise level considered, when finite-size effects are considered. For daylight, no key can be obtained when considering finite-size effects. 

While satellite-QKD could be still performed in some cases when only tip-tilt correction is performed, the use of a higher order AO system has a great impact in extending the feasibility of satellite QKD and increasing the key distribution rate.

\subsection{Background noise in DV-QKD}
\label{SI:Background}
The limiting factor for DV-QKD is given by the background noise $Y_0$, defined as the probability of having a click at Bob's side conditioned on a zero-photon pulse at Alice's side.
Two main factors contribute to $Y_0$: detector dark counts and the photons scattered from the atmosphere into the detector field-of-view.
In the simulations, two different conditions for background noise have been considered: the pessimistic case of an illuminated satellite directly reflecting sunlight into the receiving telescope and the more realistic case of a satellite flying in front of a normal daytime (nighttime) background.

For the first case, we take the photon background rate $N_b^{glonass}$, measured during a passage of one of the GLONASS satellites~\cite{Calderaro2018}, with the solar panels directly reflecting solar radiation into a $1.5$~m telescope.
We then rescale the measured $N_b^{glonass} = 1.9 \cdot 10^3$ by taking into account the different satellite distance ($400$ instead of $20000$ ~km, the different solar irradiances at the two wavelength ($532$~nm for the original article and $1550$~nm in our case) and the different interference filters ($3$~nm in the original article and $0.8$~nm in our case), and we get $N_b = 1.6 \cdot 10^5$ photons/s.
Since the intrinsic dark count rate of the detectors is $\sim 100$ photons/s, we could neglect them and estimate $Y_0$ as $N_b \Delta t$, where $\Delta t = 1$~ns is the temporal detection window, obtaining $Y_0 = 1.6 \cdot 10^{-4}$.

The estimation of the daylight sky background is made starting from the value of sky radiance for mid-summer day with $23$~km visibility, calculated with MODTRAN~\cite{Berk2014}, equal to $H_b \simeq 0.3$~W m$^{-2} \mu$m$^{-1}$sr$^{-1}$.
The measured photon background rate is calculated as
\begin{align}
    N_b = \frac{H_b \Omega_{FOV} A_{rec} B_{filter} \lambda}{h c} \eta_{opt} \eta_d,
\end{align}
where $\Omega_{FOV}$ is the field-of-view of the fiber, $A_{rec} = \pi (D_{rec}/2)^2 = 1.77$~m$^2$ is the area of the receiving telescope, $B_{filter} = 0.8$~nm is the bandwidth of the receiving filter and $\lambda = 1550$~nm is the wavelength of the radiation~\cite{gruneisen_adaptive-optics-enabled_2021}.
The efficiencies of the receiver optical system $\eta_{opt}$ and of the detectors $\eta_d$ are the same used in the simulations.
The field of view of the receiver is estimated as
\begin{align}
    \Omega_{FOV} = \pi \left( 0.45 \frac{\lambda}{D_{rec}} \right)^2 = 2.7 \cdot 10^{-12} \, sr,
\end{align}
where $D_{rec}$ is the diameter of the receiving telescope.
This estimation assumes a diffraction limited receiving system with no central obstruction~\cite{gruneisen_adaptive-optics-enabled_2021} and approximates the Airy disk as a Gaussian point spread function (PSF) with waist $0.45 \lambda f / D_{rec}$~\cite{Zhang2007}.
With these approximation, we get $N_b = 1 \cdot 10^{3}$ photons/s, from which we calculate
\begin{align}
    Y_0 = N_b \Delta t = 1 \cdot 10^{-6},
\end{align}
where $\Delta t = 1$~ns is the temporal detection window.

During the night, sky background becomes negligible with respect to the intrinsic dark counts of the detectors.
Therefore, we consider the background rate to be $N_b = 200$ photons/s, corresponding to $Y_0 = N_b \Delta t = 2 \cdot 10^{-7}$.

\subsection{Estimation of DV-QKD parameters from the properties of the channel}
\label{sec:channelProperties}
The gain $Q_{\mu}$, defined as the fraction of signal pulses giving a detection at Bob's side, can be evaluated as~\cite{vasylyev_satellite-mediated_2019}
\begin{equation}
\begin{split}
    Q_{\mu}(T^2) &= \sum_{i=0}^{\infty} Q_{\mu,i} (T^2) = \sum_{i=0}^{\infty} Y_i (T^2) P_{\mu,i} \\&= 1 - e^{-\eta_d \eta_{opt} \mu T^2} (1 - Y_0),
    \end{split}
\end{equation}
where $\eta_d$ is the detector efficiency, $\eta_{opt}$ is the overall efficiency of the optical setup at the receiver and $T^2$ is the instantaneous channel transmission efficiency.
The $i$-photon gain $Q_{\mu,i}(T^2)$ is the fraction of signal pulses giving a detection at Bob's side and coming from an $i$-photon pulse at Alice's side and can be evaluated as $Q_{\mu,i}(T^2) = Y_i(T^2) P_{\mu,i}$, where $Y_i(T^2)$ is the yield of the $i$-photon state, \emph{i.e.}, the probability of a detection at Bob's side given that Alice send an $i$-photon pulse, and $P_{\mu,i} = e^{-\mu} (\mu / i!)$ is the probability that Alice sends $i$ photons in her pulse.
The yield of the $i$-photon state is
\begin{align}
    Y_i = 1 - (1 - Y_0)(1 - \eta_d \eta_{opt} T^2)^i.
\end{align}
The $0$-photon yield $Y_0$, \emph{i.e.}, the probability of having a detection given that Alice sends no photons, is due to the combination of detector dark counts and sky-noise photons and represents the limiting factor for DV-QKD.
The probability of having an error conditioned on Alice sending $i$ photons is
\begin{align}
    e_i Y_i = e_0 Y_0 + e_d \left[ 1 - (1 - \eta_d \eta_{opt} T^2)^i \right] (1 - Y_0),
\end{align}
where $e_0 = 0.5$ is the error probability of a background photon and $e_d$ is the probability that a received photon hits the wrong detector.
From this formula, it is possible to calculate the $i$-photon error rate $E_{\mu,1} Q_{\mu,1} (T^2) = e_1 Y_1 P_{\mu,1}$ and the overall error rate
\begin{align*}
\begin{split}
    E_{\mu} Q_{\mu}(T^2) &= \sum_{i=0}^{\infty} e_i Y_i P_{\mu,i} \\&= e_0 Y_0 + e_d ( 1 - e^{-\eta_d \eta_{opt} \mu T^2} ) (1 - Y_0).
    \end{split}
\end{align*}

\subsection{Finite-key analysis of DV-QKD}
\label{sec:finiteKeyDV}
The finite key protocol studied in this work is a two-decoy efficient BB84.
It uses the basis $Z$, with probability $q$, for the key exchange and the basis $X$, with probability $1-q$, for monitoring the Eavesdropper's activities.
Signal pulses have a mean number of photons per pulse $\mu$ and are chosen with probability $p_{\mu}$, while the weak decoy state has a mean of $\nu$ photons per pulse and is chosen with probability $p_{\nu}$.
The other decoy state is the vacuum, chosen with probability $1 - p_{\mu} - p_{\nu}$.
The number of secure bit is~\cite{Lim2014,Yin2020}
\begin{align}
    l = s^L_{Z,0} + s^L_{Z,1} \left[ 1 - h(\phi^U_Z) \right] - \lambda_{EC} - 6\log_2\frac{21}{\varepsilon_{sec}} - \log_2\frac{2}{\varepsilon_{cor}},
\end{align}
where $s^L_{Z,0}$, $s^L_{Z,1}$ and $\phi^U_Z$ are, respectively, the lower bound on the number of vacuum and single-photon events and the upper bound on the phase error rate associated with single-photon events.

The lower bound on vacuum events in the $Z$ basis is given by~\cite{Lim2014}
\begin{align}
    s^L_{Z,0} = \tau_0 n^{-}_{Z,0},
\end{align}
where $\tau_n = \sum_{k \in \{\mu,\nu,0\}} e^{-k} k^n p_k / n!$ and
\begin{align}
    n^{\pm}_{Z,k} := \frac{e^k}{p_k} \left[ n_{Z,k} \pm \sqrt{\frac{n_Z}{2} \ln\frac{21}{\varepsilon_{sec}}} \right],
\end{align}
where $k \in \{\mu,\nu,0\}$ indicates the statistics used and $n_{Z,k}$ indicates the number of received photons for basis $Z$ and statistics $k$.
This value is derived by applying the Hoeffding inequality~\cite{Hoeffding1963}, with $n_Z = \sum_{k \in \{\mu,\nu,0\}} n_{Z,k}$ the total number of received events in the $Z$ basis.
The lower bound on single photon events is estimated as~\cite{Lim2014}
\begin{align}
    s^L_{Z,1} = \frac{\tau_1 \mu \left[ n^{-}_{Z,\nu} - n^{+}_{Z,0} - \frac{\nu^2}{\mu^2}\left(n^{+}_{Z,\mu} - \frac{s^L_{Z,0}}{\tau_0}\right) \right] }{\nu (\mu-\nu)}.
\end{align}
The phase error rate $\phi^U_{Z}$ is estimated by looking at the bit error rate in the mutually unbiased basis $X$, starting from
\begin{align}
    v^U_{X,1} = \tau_1 \frac{m^{+}_{X,\nu} - m^{-}_{X,0}}{\nu},
\end{align}
where
\begin{align}
    m^{\pm}_{X,k} := \frac{e^k}{p_k} \left[ m_{X,k} \pm \sqrt{\frac{m_X}{2} \ln\frac{21}{\varepsilon_{sec}}} \right],
\end{align}
for $k \in \{\mu,\nu,0\}$ is the bound in the number of errors for the $X$ basis for statistics $k$ and $m_X = \sum_{k \in \{\mu,\nu,0\}} m_{X,k}$ is the total number of errors for the $X$ basis.
The phase error rate of single-photon events in the $Z$ basis is therefore bounded by
\begin{align}
    \phi^U_Z = \frac{v^U_{X,1}}{s^L_{X,1}} + \gamma\left( \varepsilon_{sec}, \frac{v^U_{X,1}}{s^L_{X,1}}, s^L_{X,1}, s^L_{Z,1} \right),
\end{align}
with
\begin{align}
    \gamma(a,b,c,d) := \sqrt{\frac{(c+d)(1-b)b}{c d\ln2} \log_2 \left( \frac{c+d}{c d(1-b)b} \cdot \frac{21^2}{a^2} \right) }.
\end{align}
The bound $s^L_{X,1}$ is calculated in the same way as $s^L_{Z,1}$, just exchanging $Z$ with $X$ in the equation.

In the simulations, the number of events and errors are calculated using a procedure similar to the one used for the asymptotic case, as~\cite{Yin2020}
\begin{align}
    n_{b,k} = N q_b^2 p_k \mathbf{E}[Q_{k}]
\end{align}
and
\begin{align}
    m_{b,k} = N q_b^2 p_k \mathbf{E}[E_k Q_k],
\end{align}
with $b \in \{Z,X\}$ the basis and $q_b = q$ if $b = Z$ and $1-q$ if $b = X$.
As in the asymptotic limit, the values are computed as an average over channel fluctuations.
From the number of secure bits it is possible to calculate the secure key rate $K = l/N$, where $N$ is the number of exchanged symbols.\\

\end{document}